\documentclass{pasj00}
\draft

\begin{document}
\SetRunningHead{S. F. Yamada et al.}{Mahoroba11}
\Received{2005/1/dd}
\Accepted{yyyy/mm/dd}

\title{An Intermediate-band imaging survey for high-redshift Lyman Alpha Emitters: 
The Mahoroba-11}

\author{Sanae F. \textsc{Yamada},\altaffilmark{1} 
Shunji S. \textsc{Sasaki},\altaffilmark{1}  
Ryoko \textsc{Sumiya},\altaffilmark{1}  
Kazuyoshi \textsc{Umeda},\altaffilmark{1}  \\
Yasuhiro \textsc{Shioya},\altaffilmark{1}  
Masaru \textsc{Ajiki}\altaffilmark{1} , 
Tohru \textsc{Nagao},\altaffilmark{2,3}  \\
Takashi \textsc{Murayama},\altaffilmark{1}  
and 
Yoshiaki \textsc{Taniguchi}\altaffilmark{1} 
}


\email{shioya@astr.tohoku.ac.jp}
\email{tani@astr.tohoku.ac.jp}

\altaffiltext{1}{Astronomical Institute, Graduate School of Science, Tohoku University, \\
                 Aramaki, Aoba, Sendai 980-8578}
\altaffiltext{2}{National Astronomical Observatory of Japan,\\
                 2-21-1 Osawa, Mitaka, Tokyo 181-8588}
\altaffiltext{3}{INAF --- Osservatorio Astrofisico di Arcetri,\\
                 Largo Enrico Fermi 5, 50125 Firenze, Italy}


%


\maketitle

\begin{abstract}
    We present results of our intermediate-band optical imaging
 survey for high-$z$ Ly$\alpha$ emitters (LAEs) using the prime focus
 camera, Suprime-Cam, on the 8.2m Subaru Telescope. In our survey,
 we use eleven filters; four broad-band filters ($B$, $R_{\rm c}$,
 $i^\prime$, and $z^\prime$) and seven intermediate-band filters
 covering from 500 nm to 720 nm; we call this imaging program as
 the Mahoroba-11. The seven intermediate-band filters are selected
 from the IA filter series that is the Suprime-Cam intermediate-band filter system 
 whose spectral resolution is $R = 23$. Our survey has been
 made in a $34^\prime \times 27^\prime$ sky area in the Subaru
 XMM Newton Deep Survey field.
    We have found 409 IA-excess objects that provide us a large
photometric sample of strong emission-line objects. 
Applying the photometric redshift method to this sample,
we obtained a new sample of 198 LAE candidates at $3 < z < 5$. 
We found that there is no evidence for evolution of the number density and 
the star formation rate density for LAEs with $\log L({\rm Ly}\alpha) 
({\rm erg \; s^{-1}}) > 42.67$ between $z \sim 3$ and 5. 

\end{abstract}

\section{Introduction}

Formation and evolution of galaxies have been intensively discussed
in this decade. This progress has been supported by the Hubble Space 
Telescope (HST) and ground-based 8-10 m class telescopes such as the W. M.
Keck telescopes, VLT, the Gemini Telescopes, and the Subaru Telescope. In particular, the Hubble
Deep Fields (Williams et al. 1996, 2000) and the Hubble Ultra Deep Field 
(Beckwith et al. 2003) make it possible 
to search for very faint galaxies  beyond $z = 5$ (e.g., Lanzetta et al.
1996; Weymann et al. 1998; Bunker et al. 2004: Yan, Windhorst 2004).

Early star formation in galaxies could occur beyond $z = 5$ 
(e.g., see for reviews, Taniguchi et al. 2003b, Spinrad 2003). 
Yet, the peak of star formation in the universe
appears to occur at $z \sim 1$ (Madau et al. 1996; Steidel et al. 1999;
Ouchi et al. 2004; Giavalisco et al. 2004). 
Although this observational trend has been 
widely accepted, we have not yet fully understood which physical
processes play crucially important roles during the course of
evolution of galaxies.
Therefore, in order to understand the whole history of galaxies 
in the universe from high redshift to the present day, it is important 
to carry out systematic searches for galaxies at any redshift. 

It has been also claimed that the basic dynamical properties of galaxies
could be determined at $1 < z < 3$; i.e., the origin of the Hubble type
of galaxies could be established at this redshift interval 
(e.g., Kajisawa, Yamada 2001).
In addition, early star formation in galaxies as well as early 
growth of supermassive black holes in their nuclei should be 
clarified unambiguously in terms of the hierarchical structure
formation paradigm (Madau et al. 1996; Barger et al. 2001;
Kauffmann, Haehnelt 2000). 

In order to explore this issue, it is absolutely necessary to
investigate  observational properties of galaxies beyond $z=1$.
Indeed, many observational programs have been devoted to
this problem by finding high-$z$ galaxies. Searches for such
high-$z$ galaxies have been carried out mainly by the following two methods.
One is the optical broad-band color selection technique (e.g., Steidel,
Hamilton 1992; Steidel et al. 1999; Madau et al. 1996; Lanzetta 
et al. 1996; Giavalisco et al. 2004 and references therein). 
The other method is deep searches for strong Ly$\alpha$ emitters
by using an optical narrow-band filter (Hu et al. 1996; Cowie, Hu 1998;
Kudritzki et al. 2000; Rhoads et al. 2000; Ajiki et al. 2003;
Kodaira et al. 2003; Taniguchi et al. 2005). 
We also note that the slitless grism spectroscopy is another 
method to search for the high-z galaxies (Kurk et al. 2004; Martin \& Sawicki 
2004; Pirzkal et al. 2004). 

The first method provides a relatively large sample of high-$z$ 
Lyman break galaxies (LBGs) and their follow-up spectroscopy tells us
global properties of high-$z$ galaxies. 
However, such studies are biased to investigations of relatively
bright galaxies because a certain magnitude-limited sample is 
used in such a study. On the other hand, the second method
provides us a sample of strong emission-line objects, most of 
which are missed in a magnitude-limited sample, 
although it is not clear whether or not they are typical faint normal galaxies.
However, because of the narrowness of the filter bandpass,
such a study can only probe strong emission-line objects 
in a small volume of the universe (see for a review,
Taniguchi et al. 2003b). 

In order to overcome the demerit of searches with a
narrow-band filter, searches with multi narrow-band filters
have been made in some cases (Rhoads et al. 2000;
Shimasaku et al. 2004). Even in their pioneering surveys,
only a couple of filters were used and thus their survey 
volumes were still small. More recently, new surveys with 
a combination between typical broad-band filters and 
intermediate- and narrow-band filters have been conducted.
The most successful survey made so far is the COMBO-17 survey in which 
17 filters are used (Wolf et al. 2003a; see also Bell et al.
2004); 5 broad-band filters and 12 medium-band ones. 
Indeed, this survey has been used to make a systematic search
for strong emission-line objects such as  active galactic nuclei
between $z \sim 1$ and $z \sim 5$ (Wolf et al. 2003b).   

In order to search for strong Ly$\alpha$ emitters systematically,
we have conducted a new deep optical imaging survey with 11 filters
using the prime focus camera, Suprime-Cam (Miyazaki et al. 2002), on 
the 8.2m Subaru Telescope (Kaifu et al. 2000; Iye et al. 2004).
In this survey, we use a set of seven
intermediate-band (IA) filters covering wavelengths from $\approx$
527nm to 709nm, corresponding to the Ly$\alpha$ redshift from
$z=3.4$ and $z=4.8$ (Hayashino et al. 2000; Taniguchi 2004;
Fujita et al. 2003; Taniguchi et al. 2003b;  Ajiki et al. 2004). 
This new data set allows us to investigate
the cosmic star formation history systematically from $z=4.8$
to $z=3.3$ for the first time. In this paper, we present our results
and then discuss the nature of strong Ly$\alpha$ emitters at
high redshift.

Throughout this paper, we adopt a flat universe with $\Omega_{\rm m}=0.3$, 
$\Omega_{\Lambda}=0.7$, and 
$H_0=70{\hspace{0.8mm}} {\rm {\hspace{0.8mm}}km \; s^{-1} Mpc^{-1}}$, 
and we use the AB magnitude system (e.g. Oke 1974). 

\section{Observations and data reduction}

\subsection{Observations}

Deep and wide-field $B$-, $R$-, $i'$-, and $z'$-band imaging data 
of a southward $30'\times 24'$ area in the 
Subaru/XMM-Newton Deep Survey Field (the SXDS field\footnote{
See http://www.naoj.org/Science/SubaruProject/SDS/.}) centered at
$\alpha$(J2000) = $2^{\rm h} 18^{\rm m} 00^{\rm s}$ and
$\delta$(J2000) = $-5^\circ 12 ' 00''$ were obtained using 
the Suprime-Cam (Miyazaki et al. 2002) 
on the 8.2m Subaru Telescope during a period between 2000 August, 
and 2002 January by the SXDS team. 
The Suprime-Cam consists of ten CCD chips, each of which has 2048
 $\times$ 4096 pixels, providing a field of view of $27^\prime
\times 34^\prime$. 

In the SXDS field, X-ray deep survey by the XMM-Newton
satellite as well as 
multi-broad-band imaging observations are on-going over 1.3 square degree area. 
Therefore, the field is one of the most suitable target fields
for our IA filter survey. 
Carefully examining the entire SXDS field, 
we selected the southward field of SXDS field 
which do not contain any bright stars. 
Note that a sky area of 13.$^\prime 7$ $\times 13.^\prime 7$ 
in the central part of the SXDS field was already observed 
by Fujita et al. (2003) using an IA filter, $IA574$. However,
our observing field does not overlap with it (see figure 2). 

In addition to these broad-band image data,  
intermediate-band images were obtained during an open-use observing
program, S02B-163 (PI = K. Kodaira),
using the following IA filters: $IA527$, $IA574$, $IA598$, 
$IA624$, $IA651$, $IA679$, and $IA709$, on Suprime-Cam / Subaru Telescope 
during a period between 2002 October and 2002 November.
The center wavelength and $FWHM$ of these IA filters are summarized 
in table 1; see also \verb|http://www.awa.tohoku.ac.jp/~tamura/astro/filter.html |, 
Hayashino et al. (2000, 2003), and Taniguchi (2001).
The transmission curves of the filters 
used in our observations are shown in figure 1.
A journal of our all observations is given in table 1.
All of the observations were made under photometric conditions, 
and the seeing size was $\simeq 1.''0$. 

\subsection{Data Reduction}

The individual CCD data were reduced and
combined using IRAF and the mosaic-CCD data reduction software 
developed by Yagi et al. (2002).
The following spectrophotometric standard stars were observed to 
calibrate the IA filter imaging data; 
LDS749B, LTT9491, G93-48, and Feige110 [references here Oke (1974), Oke (1990), 
Landolt (1992), and Stone (1996)].
The combined images for the individual bands were aligned and 
smoothed with Gaussian kernels to match their seeing sizes. 
The final images cover a contiguous
944 arcmin$^2$ area ($34'.71$ $\times$ $27'.20$) with a PSF FWHM of $1.''04$ 
for the broad-band and IA-band data.
The final IA image stacked with the seven IA images is shown in figure 3. 
After masking out areas which are affected by the starlight,
we obtained the final image whose effective sky coverage is
526 arcmin$^2$.
Note that the data reduction was made by the team of S02B-163 program 
in which the authors in this paper were included as collaborators.  

\section{Results}

\subsection{Source Detection and Photometry}

For photometry and source detection of our observational data, 
we use SExtractor version 2.1.7 (Bertin \& Arnouts 1996). 
As for the source detection in all images,
we use the limiting magnitudes for a 3$\sigma$ detection with a
$2''$ diameter aperture : $B=28.2$, $R_{\rm c}=27.4$, $i'=27.0$, $z'=25.8$, 
$IA527=26.8$, $IA574=26.5$, $IA598=26.5$, $IA624=26.7$, 
$IA651=26.7$, $IA679=26.8$, and $IA709=26.6$.
In the above source detection, we have detected $\sim 70000$ sources
down to $IA = 27$ in each IA image.

As shown in the left panel of figure 4, we can detect objects to $\sim 27$ mag 
with above condition. 
In order to examine the completeness of our IA images,
we show results of the number counts in the IA imaging as
a function of AB magnitude in figure 4.
These results show that our IA imaging appears complete 
down to 26.5 -- 27 mag for each IA image.

In order to examine further how accurately we select emission-line objects
in our IA images, we have made simulations using IRAF ARTDATA
(e.g., Kajisawa et al. 2000; Fujita et al. 2003). 
For this purpose, we generate two sets of 300 model galaxies
for each magnitude interval, $\Delta m=0.2$ mag. 
Model galaxies in the first set 
obey the de Vaucouleurs $r^{1/4}$ law light distribution
while those in the second set obey the exponential law.
Their sky positions, half-light radius (from 1 to 7 kpc),
and ellipticities are randomly determined. We put these galaxies 
into the CCD data 
together with Poisson noises. 
After smoothing model-galaxy images to match to the seeing size 
of our observation, 
we try to detect them using SExtractor with the same procedure 
as that we used in our reduction. 
In the right panel of figure 4, 
the detectability is shown as a function of IA magnitude
for each IA filter. 
These results also show that our IA imaging appears complete 
down to 26.5 -- 27 mag for each IA image.

\subsection{Selection of IA-Excess Objects}

We describe our procedures to select IA-excess objects from our data set. 
For this purpose, we need a continuum image for each IA image. 
Since the central wavelength of each IA filter does not generally match to 
that of a certain broad band filter, we have to generate a custom continuum 
image for each IA filter. 
In previous studies (e.g., Steidel et al. 2000; Fujita et al. 2003), 
the broad band image whose effective wavelength is close to that 
of the concerned narrow-band or intermediate-band one was used as a continuum image.
However, if we follow this way, continuum break objects such as LBGs 
could also be selected as a strong IA-excess object. 
In order to reject such objects, 
we have adopted the image of the broad filter band ($D[IA]$) 
whose central wavelength is longer than that of the concerned IA filter 
as a continuum image: $D[IA]$ is 
$R_{\rm c}$-band for $IA527$, $IA574$, and $IA598$, 
and $i'$-band for $IA624$, $IA651$, $IA679$, and $IA709$.
Hereafter, we refer these continuum  $D$[$IA$],
or $D$[$IAnnn$] where $nnn$ = 527, 574, 598, 624, 651, 679, or 709. 

In figure 5, we show the color - magnitude diagram between $D[IA]-IA$ and $IA$ 
for objects in our $IA$ catalogs. 
Taking the scatter in the $D[IA]-IA$ color into account, we define
strong IA-excess objects with the following criterion, 
\begin{equation}
D[IA]-IA ~ \geq ~ mag(EW_{\rm rest}=20{\rm \AA~}), 
\end{equation}
where $mag (EW_{\rm rest}=20$\AA~) is the magnitude corresponding to $EW_{\rm rest}=20$\AA ~ determined as,
\begin{equation}
D[IA] - IA = -2.5 {\rm log} \frac{\it FWHM(IA)}{EW_{\rm obs} + FWHM(IA)},
\end{equation}
where $FWHM(IA)$ is the $FWHM$ of IA filter (see table 1). 
$mag$($EW_{\rm rest}$=20\AA~) are
0.332, 0.325, 0.314, 0.320, 0.311, 0.311, and 0.339 
for $IA527$, $IA574$, $IA598$, $IA624$, $IA651$, $IA679$, and $IA709$, respectively. 
These criteria are shown by horizon lines in figure 5.
We also add the criteria for error,
\begin{equation}
D[IA]-IA ~ > ~ 3\sigma(D[IA]-IA) \hspace{5mm}{\rm and}\hspace{5mm} IA ~ < ~ 3\sigma(IA) \\
\end{equation}
(see curved lines in figure 5).
Then, we select 74, 47, 42, 48, 60, 77, and 61 strong IA-excess objects 
in the $IA527$, $IA574$, $IA598$, $IA624$, $IA651$, $IA679$, 
and $IA709$ filter images, respectively.

\subsection{Selection of Lyman $\alpha$ Emitter Candidates}

Because of the large observed equivalent width, 
it seems that most IA-excess objects may be LAEs at high redshift. 
However, there may be still a possibility that some of them are 
strong emission-line objects at lower redshifts; e.g., 
H$\alpha$ emitters at $z \sim 0.00$ to 0.10, 
[O {\small III}] $\lambda$ 5007 emitters at $z \sim 0.03$ to 0.45, and 
[O {\small II}] $\lambda$ 3727 emitters at $z \sim 0.38$ to 0.94. 
Here, we try to select LAEs from IA-excess objects. 
Although a color - color diagram is usually used to separata the LAEs from low-z 
emission-line galaxies (e.g., Taniguchi et al. 2005), 
it seems difficult to separate LAEs properly especially for blue IA bands. 
We therefore adopt the photometric redshift (SED fitting) technique 
using all 11 bands photometric data. 
Using this method, we can select LAEs with enough accuracy for all IA bands. 
The detailed discussion on the accuracy of the photometric redshift is shown 
in appendix 1. 


For this purpose, we generate model galaxy SEDs using  the population
 synthesis model, GALAXEV, developed by Bruzual \& Charlot (2003).
The SEDs of local galaxies are well reproduced 
by models whose star-formation rate declines exponentially 
(the $\tau$ model) : i.e., $SFR(t) \propto {\rm exp}(-t/\tau)$, where $t$ is the
age of galaxy and $\tau$ is the time scale of star formation.
 In this study, we use $\tau=1$ Gyr models with Salpeter's initial mass function
(the power index of $x=1.35$ and the stellar mass range of 
$0.1 \leq m/M_{\odot} \leq 100$) to derive various SED types.
We note that the SED templates derived by Coleman et al. (1980) for  
elliptical galaxies, Sbc, Scd, Irr, and SB, correspond to those using 
$t=$8, 4, 3, 2, and 1 Gyr, respectively.
We calculate SEDs with ages of t=0.1, 0.5, 1, 2, 3, 4, and 8 Gyr.

When we use only optical broad band photometry for estimates of the
photometric redshift, we need not seriously take account of 
the contribution of emission lines to the broad band flux. 
However, in our case, we cannot neglect the contribution 
of some strong emission lines because our photometric data contain
intermediate-band data. 
In order to include possible emission-line fluxes into our models,
we calculate the number of ionizing photons, $N_{\rm Lyc}$, 
from the SED which is calculated from the above population synthesis model. 
Then we can estimate H$\beta$ luminosity, $L$(H$\beta$) 
using the following formula (Leitherer \& Heckman 1995), 

\begin{equation}
L(H\beta) = 4.76 \times 10^{-13} N_{\rm Lyc}\hspace{3mm} \rm erg \hspace{1mm} s^{-1}.
\end{equation}
Other strong emission-line luminosities, such as for [O {\sc ii}], 
[O {\sc iii}], H$\alpha$, and so on, are estimated by typical line ratios
relative to H$\beta$ (PEGASE: Fioc \& Rocca-Volmerange 1997).

In addition to the contribution of strong emission lines to SEDs,
we also take the following two effects into account in SED templates.
One is the absorption by interstellar medium in a galaxy itself.
For this correction, we use the starburst reddening curve 
of Calzetti et al. (2000), 
\begin{equation}
F_i(\lambda) = F_o (\lambda) 10^{0.4E_s(B-V)k'(\lambda)}, 
\end{equation}
where $F_i(\lambda)$ is the intrinsic stellar continuum, 
$F_o(\lambda)$ is the observed stellar continuum, 
and $k^\prime(\lambda)=A^\prime(\lambda)/E_s(B-V)$ is the starburst reddening curve. 
The expression of $k^\prime(\lambda)$ is 

\begin{eqnarray}
k^\prime(\lambda) & = & 2.659(-1.857+1.040/\lambda) + R^\prime_V 
 ~ \; \; \; {\rm for} ~ 0.63 \mu{\rm m} \le \lambda \le 2.20 \mu{\rm m}; \nonumber \\
 & = & 2.659 (-2.156 -1.509/\lambda -0.198/\lambda^2 + 0.011/\lambda^3 + R^\prime_V) \nonumber \\
 &   & ~ {\rm for} ~ 0.12 \mu{\rm m} \le \lambda < 0.63 \mu{\rm m}, 
\end{eqnarray}
where $R^\prime_V = A^\prime(V)/E_s(B-V) = 4.05$. 
And another effect is the absorption from intergalactic gas. 
Following the method in Madau et al. (1996), we describe the observed mean 
specific flux of a source at redshift $z_{\rm em}$ as 
\begin{equation}
\langle f(\nu_{\rm obs})\rangle = 
\frac{(1+z_{\rm em})L(\nu_{\rm em})}{4\pi d^2_{L}}\langle e^{-\tau} \rangle, 
\end{equation}
where $\nu_{\rm obs}=\nu_{\rm em}/(1 + z_{\rm em})$, $d_L$ is the luminosity distance
corresponding to $z_{\rm em}$ 
and $\langle e^{-\tau} \rangle$ is the average transmission. 

Then, we try to fit the observed SEDs of IA-excess objects 
using SED templates generated by the above procedures. 
In this fitting, we calculate the standardized ``relative likelihood'' 
between $z_{\rm ph}$=0 and $z_{\rm ph}$=5.0.
We adopt a redshift of the primary peak of likelihood as the
most probable photometric redshift
only when the peak value is higher by a factor of two than that of the
secondary peak. Otherwise, we judge that we cannot obtain any
reliable photometric redshift.
In this way, we have obtained photometric redshifts for 
64, 40, 20, 26, 16, 17, and 15 LAEs found 
in IA527, IA574, IA598, IA624, IA651, IA679, and IA709 
selected catalogs, respectively. 
We summarize the number of LAEs in table 2. 
The photometric catalogs of LAEs are shown in table 3. 

We show the spatial distributions of LAEs in figure 6. 
The areas shown by grey color are the masked region.
Although it is interesting to investigate clustering properties
of these objects, a significant part of the sky area is masked
on each IA image and thus it seems difficult to obtain
a firm result on the clustering properties.
Therefore, we do not discuss this issue further in this paper.

\subsection{Equivalent Widths}

We show the distributions of observed equivalent widths, 
$EW_{\rm obs}$, for the high-$z$ LAE sample found in 
the previous section for those found in each IA filter
in figure 7.
We note that a LAE found in $IA574$ catalog have very large $EW_{\rm obs}$;
6840.6 \AA \footnote{We comment on this large $EW_{\rm obs}$. 
The $R_{\rm c}$-band flux of this object is smaller than $2 \sigma$. 
Using the $2 \sigma$ flux, we derive the lower limit of $EW_{\rm obs} = 1288$ \AA.}. 
These data are not shown in figure 7.
We find that LAEs with smaller $EW_{\rm obs}$ tend to be more populous.
We also show the histograms for the total sample in the 
lower-right panel of figure 7.
Next, we show the distributions of rest-frame equivalent widths
for the same samples in figure 8 
where $EW_{\rm 0}$(Ly$\alpha$) = $EW_{\rm obs}$(Ly$\alpha)/(1+z)$. 

\subsection{Lyman $\alpha$ luminosity}

We estimate the Ly$\alpha$ luminosity  for each LAE
found in our survey using the following relation,
\begin{equation}
L ({\rm Ly}\alpha) = 4\pi d_L^2 f({\rm Ly}\alpha) \hspace{5mm} {\rm erg \hspace{1mm} s^{-1}},
\end{equation}
where $d_L$ is the luminosity distance, and $f$(Ly$\alpha$) is the Ly$\alpha$ flux
estimated using $D$[$IA$] and $IA$ magnitude below, 
\begin{equation}
f({\rm Ly}\alpha) = \{ f_\nu (IA) - f_\nu(D[IA]) \} \Delta \nu \hspace{5mm} {\rm erg \; s^{-1} \; cm^{-2} }, 
\end{equation}
where $f_\nu(IA) = 10^{-0.4(IA + 48.6)} \; {\rm erg \; s^{-1} \; cm^{-2} \; Hz^{-1}}$, 
$f_\nu(D[IA]) = 10^{-0.4(D[IA] + 48.6)} \; {\rm erg \; s^{-1} \; cm^{-2} \; Hz^{-1}}$, 
and $\Delta \nu$ is the bandwidth in unit of Hz. 
In this calculation, we assume all the LAEs exist at the redshift shown in table 2 
for each IA-band. 

\section{Discussion}

\subsection{Number Densities of Lyman$\alpha$ Emitters }

Since we have found LAEs at $z \simeq$  3.3 -- 4.8,
we can investigate the number density evolution of LAEs
as a function of redshift. 
The faintest $\log L({\rm Ly \alpha})$ is different for different IA band: 
42.32 for IA527, 42.51 for IA574, 42.56 for IA598, 42.56 for IA624, 42.61 for IA651, 
42.63 for IA679, and 42.67 for IA709, where $L$(Ly$\alpha$) is in units 
of erg s$^{-1}$. 
In order to compare with samples which is obtained 
by the same selection limit, we use the LAEs 
which is brighter than the IA limit, 
${\rm log} L({\rm Ly}\alpha) = 42.67$. 
The number of LAEs brighter than ${\rm log} L({\rm Ly}\alpha) = 42.67$ 
is 10 (IA527), 15 (IA574), 10 (IA598), 15 (IA624), 12 (IA651), 15 (IA679) 
and 15 (IA709), respectively,  where $L$(Ly$\alpha$) is again in units 
of erg s$^{-1}$. 
Our effective survey area is 526 ${\rm arcmin^2}$.
Then the volume covered by each filter is 
3.36$\times10^5$ (IA527), 
3.63$\times10^5$ (IA574), 
3.86$\times10^5$ (IA598), 
3.85$\times10^5$ (IA624), 
4.05$\times10^5$ (IA651), 
4.12$\times10^5$ (IA679), 
and 3.78$\times10^5$ (IA709)
 ${\rm Mpc^{3}}$, respectively.
Therefore, we obtain the number density of LAEs for each
filter as follows; $n$(LAE) $\simeq$
$3.0 \times 10^{-5}$ Mpc$^{-1}$ at $z \simeq 3.34$, 
$4.1 \times 10^{-5}$ Mpc$^{-1}$ at $z \simeq 3.72$, 
$2.6 \times 10^{-5}$ Mpc$^{-1}$ at $z \simeq 3.93$, 
$3.9 \times 10^{-5}$ Mpc$^{-1}$ at $z \simeq 4.12$, 
$3.0 \times 10^{-5}$ Mpc$^{-1}$ at $z \simeq 4.35$, 
$3.6 \times 10^{-5}$ Mpc$^{-1}$ at $z \simeq 4.58$, and 
$3.9 \times 10^{-5}$ Mpc$^{-1}$ at $z \simeq 4.82$. 
These results are shown in figure 9. 
Taking the estimated error into account, 
we may conclude that the LAE number density is constant at $3.3 < z < 4.8$. 

Let us compare our results with
published results. Fujita et al. (2003) obtained 
$n$(LAE) $\simeq 6.4 \times 10^{-5}$ at $z \simeq 3.7$,
being smaller by a factor of two than our value at $z \simeq 3.7$.
This may be attributed to their survey is shallower than ours.
Cowie \& Hu (1998) made deep imaging survey for LAEs at $z \sim 3.4$ 
in the Hubble Deep Field North and SSA22 and then found
$n$(LAE) $\sim 1 \times 10^{-3}$ Mpc$^{-1}$. Kudritzki et al. (2000) obtained 
$n$(LAE) $\sim 1 \times 10^{-3}$ Mpc$^{-1}$ for a blank field at $z \simeq 3.1$.
Steidel et al. (2000) also obtained a large value for the SSA22a field; 
$n$(LAE) $\sim 4 \times 10^{-3}$ Mpc$^{-1}$ at $z \simeq 3.1$. Although 
the field studied by Steidel et al. (2000) is a so-called 
over density region, the LAE number density appear to depend on
the survey depth. In this respect, since our LAE survey is a homogeneous
one for LAEs between $z \sim$ 3.3 - 4.8, the results shown
in figure 9 can be used to investigate the evolution of 
LAE number density for the first time. Therefore, the constant 
number density of LAEs is one of important results in this study. 

\subsection{Ly$\alpha$ Luminosity Distribution}

In order to estimate contribution of LAEs 
to the cosmic star formation rate density at high redshift, 
we need a reliable Ly$\alpha$ luminosity function
based on a large sample of LAEs.
Since we have found 198 reliable candidates of LAEs
at $z \simeq$ 3.3 -- 4.8, our sample is useful for this purpose.

In figure 10, we show the distribution of Ly$\alpha$ 
luminosities. We note that in this figure each bin width is 
$\log L({\rm Ly}\alpha)=0.2$ where $ L({\rm Ly}\alpha)$ is 
in units of erg s$^{-1}$. 
Our derived Ly$\alpha$ luminosities range from $10^{42.3}$ to $10^{43.2}$ 
erg $\rm s^{-1}$. These luminosities are higher than typical ones
found in previous LAE surveys with use of a narrowband filter.
The reason for this is that our IA filters have wider FWHMs than
the narrowband filters used in the previous LAE surveys and thus
our survey tend to find LAEs with a larger Ly$\alpha$ equivalent width.
Therefore, we may underestimate the density at low luminosity range ($\log L({\rm Ly}\alpha) < 42.5$) 
by the limit of $L$(Ly$\alpha$) at high-redshift. 
. 

\subsection{UV Luminosity Distribution}

Since the LAE candidates are selected by the estimation of 
photometric redshift, 
their $UV$ magnitudes are bright enough  
to be detected on our $R$ or $i^{\prime}$ image. 
Therefore, these candidates would be detected as LBG candidates. 
In this respect,  our LAE candidates can be regarded as subsamples of LBGs. 
Therefore, it is interesting to compare UV luminosities of our LAEs
with those of LBGs at similar redshifts.
For this purpose, we estimate UV luminosities of our LAEs
using broad band photometric data.
We adopt $R_{\rm c}$-band or $i'$-band as UV continuum; 
$R$ for the LAEs found in $IA527$, $IA574$, and $IA598$
catalogs, while $i'$ for those in
 $IA624$, $IA651$, $IA679$, and $IA709$ catalogs. 

The results for LAEs of each IA catalog are shown in figure 11.
In the lower-right panel of figure 11, we also show 
the results for all LAEs in our study.
In this panel, we show the results of $z\simeq 5.7$ LAE survey made 
by Hu et al. (2004) ({\sl filled squares}).

Our results appear to be quite similar to their results although
our redshift range ($3.3 < z < 4.8$) is smaller than the redshift 
at the survey field of Hu et al.  
This suggests that the UV luminosity of LAEs
does not show strong evolution from $z \simeq 5.7$ to
$z \simeq 3.3$ although we need more data to confirm this.

Then we compare our results with those obtained for
LBGs at $z \sim$ 3 -- 4 (Steidel et al. 1999). 
As shown in figure 11,
the number density of LBGs is systematically higher than that of LAEs
in any UV luminosity (by a factor $\sim 5$).
This may in part explain why the SFRD derived
from LAEs is significantly smaller than that derived from LBGs.
This will be discussed later (see Section 4.4.3).

\subsection{Star Formation Rate}

\subsubsection{Star Formation Rate Based on Ly$\alpha$ Luminosity}

We estimate the SFR for the emitters 
at each seven IA filters. 
We adopted the formula from Kennicutt (1998),

\begin{equation}
SFR = 7.9 \times 10^{-42} L{\rm (H\alpha)} \,\, M_{\odot}\,{\rm yr^{-1}}\, 
\end{equation}
where $L$(H$\alpha$) is in units of erg s$^{-1}$, 
and from the case B recombination theory (Brocklehurst 1971), 

\begin{equation}
L{\rm (Ly\alpha)} = 8.7 L{\rm (H\alpha)}, \\
\end{equation}
where $L$(Ly$\alpha$) is also in units of erg s$^{-1}$, 
we can obtain the conversion relation between the SFR 
and the Ly$\alpha$ luminosity,

\begin{equation}
SFR({\rm Ly}\alpha) = 9.1 \times 10^{-43} L{\rm (Ly\alpha)} \, M_{\odot}\,{\rm yr^{-1}}\\. 
\end{equation}
Using this relation, we estimate the SFR for all LAEs.
The results are given in figure 12.
The obtained SFR ranges between 1.8 $M_\odot$ yr$^{-1}$
and 17.0 $M_\odot$ yr$^{-1}$ with a median SFR of 
4.1  $M_\odot$ yr$^{-1}$. These results are consistent with
previous surveys for high-$z$ LAEs (e.g., Cowie \& Hu 1998;
Keel et al. 1999; Kudritzki et al. 2000; Fujita et al. 2003;
Ajiki et al. 2003; Cuby et al. 2003; Taniguchi et al. 2005).

\subsubsection{Star Formation Rate Based on UV Luminosity}

We can also estimate SFR from rest-frame UV continuum
and then we examine whether or not the SFR derived
from the Ly$\alpha$ luminosity is 
consistent with that derived from the UV continuum luminosity 
for our sample. 
As described in section 4.3, we can estimate the rest-frame
UV luminosity for all LAEs. Then we obtain the SFR based on UV
luminosity using the following relation,

\begin{equation}
SFR{\rm(UV)} = 1.4 \times 10^{-28} L_{\nu} \, {\rm M_{\odot}/yr}, \\
\end{equation}
where $L_{\nu}$ is in units of erg s$^{-1}$ Hz$^{-1}$
(Kennicutt 1998).

The use of continuum magnitudes to estimate the SFR avoids 
the extremely complex problems of both the Ly$\alpha$ escape process 
and the uncertainties in the correction of the Ly$\alpha$ fluxes 
for intergalactic scattering which are present in the determination 
of the Ly$\alpha$ luminosity function. 

\subsubsection{Comparison between $SFR$(Ly$\alpha$) with $SFR$(UV)}

Now we compare $SFR$(Ly$\alpha$) with $SFR$(UV).
The results are shown in figure 12 for LAEs found
in each IA catalog. In the final panel of this figure, 
we also show results for all LAEs.
We find that the two kinds of SFRs appears to be
consistent within a factor of two for most of our LAEs,
in particular for LAEs at redshift between $z \simeq 3.3$ (IA527) 
and $z \simeq 3.9$ (IA598). However, for most LAEs at redshift between
$z \simeq 4.1$ and $z \simeq 4.8$ (i.e., $z > 4$),
there appears a tendency that  $SFR$(Ly$\alpha$) is systematically
smaller by a factor of $\approx$ 1.3 than $SFR$(UV).
Such tendency is often found LAEs beyond $z = 5$
(e.g., Hu et al. 2004; Ajiki et al. 2003; Taniguchi et al. 2005
 and references therein).
In order to see this tendency more clearly, we show
the average $SFR$(Ly$\alpha$)/$SFR$(UV) ratio,
$\langle \log SFR({\rm Ly}\alpha)/SFR(UV) \rangle$,
as a function of redshift in figure 13;
the averages $\log SFR({\rm Ly}\alpha)/SFR(UV)$ are 0.03, 0.09, 0.92, 
$-$0.10, $-$0.04, $-$0.07, and $-$0.11 
for $IA527$, $IA574$, $IA598$, $IA624$, $IA651$, $IA679$, and $IA709$,
respectively. 
Although these averages seem to slightly decrease with 
increasing redshift as shown in figure 13, 
it can be said that they are almost constant, i.e.,
$\langle \log SFR({\rm Ly}\alpha)/SFR(UV) \rangle \simeq 1$.
We summarize the observational properties of LAEs written above in table 4. 

\subsubsection{Cosmic Evolution of the Star Formation Rate Density}

The cosmic star formation history is one of the important key issues
related to the formation and evolution of galaxies. 
In particular, the cosmic star formation history at high redshift
has been mainly investigated by using galaxy samples
based on the optical broad-band color selection; i.e.,
the Lyman break method (e.g., Madau et al. 1996; Steidel et al. 1999).
More recently, LAEs have been also used to investigate
the cosmic star formation history (e.g., Ajiki et al. 2003;
Taniguchi et al. 2005; see for a review Taniguchi et al. 2003b).
The use of LAE samples as well as LBG ones
is important because such strong LAEs
could contribute to the cosmic star formation rate significantly.
However, the majority of them may be missed in broad-band
photometric samples because they are often too faint
to be contained in a broad-band
magnitude-limited sample.
One concern with surveys with a narrowband filter is that
the redshift coverage is inevitably small. Therefore, 
information on the SFR is quite limited 
for a certain, discrete redshift.
Since the LAEs found in this study
are located from $z \simeq 4.8$ to $z \simeq 3.3$, they allow
us to investigate the cosmic star formation history viewed
from LAEs systematically in the concerned redshift range.
This is indeed one of main purposes of our IA filter survey.

We obtain the star formation rate density (SFRD) for
each LAE sample at $z \sim 3.3$ -- 4.8. 
The results are shown in the upper panel of figure 14 with previous results. 
Here, we use SFR calculated from Ly$\alpha$ emission. 
The SFRDs shown in figure 14 are the simple sum of SFRs divided by the survey volume 
for all LAE candidates (open circles in the both panels) and LAE candidates of 
$\log L({\rm Ly}\alpha > 42.67)$ (filled cirles in the lower panel). 
The errors of the SFRD which arise from the photometric error is 8 \% at most. 
Since we have not corrected for extinction for Ly$\alpha$ emission
and we have not made summing up by using a Ly$\alpha$ luminosity function, 
these SFRDs are lower limit. 

Our results show that SFRD for all LAE candidates basically decreases with increasing
redshift. 
The decreasing trend may be caused from our selection limit. 
We compare the SFRD for LAEs with $\log L ({\rm Ly \alpha}) > 42.67$ 
(the filled circles in the lower panel of figure 14). 
The SFRD derived for LAEs with $\log L ({\rm Ly \alpha}) > 42.67$ is 
nearly constant at $z=3.3$--4.8. 
We can detect the LAE candidates from log$L$(Ly$\alpha$)$\simeq 42.4$ 
to log$L$(Ly$\alpha$)$\simeq 43.5$ (see figure 10). 
However since we observe only the bright-end of the number densities, 
we may miss a number of low-luminosity LAEs.

The data points with an upward arrow in figure 14 show the results 
of LAE searches. 
Since Ly$\alpha$ emission from high-redshift objects is absorbed 
either by the intergalactic medium  or by the gas in the system itself or both,
the blueward flux of Ly$\alpha$ emission should be 
underestimated. 
Therefore it could be desired that 
the SFRD using Ly$\alpha$ method must be corrected for this effect.

\section{Summary}

We have presented seven optical intermediate-band 
and multicolor observations of the Subaru / XMM-Newton Deep Field
obtained with the
Suprime-Cam on the 8.2 m Subaru telescope. 
The intermediate-band image covered a sky area of 
526 ${\rm arcmin^{2}}$ in the Subaru/XMM-Newton Deep Field (Ouchi et al. 2001). 
Our survey volume amounts to 2.9$\times 10^{6} {\rm Mpc^{3}}$ 
when we adopt a flat universe with $\Omega_{\rm matter}$ = 0.3, 
$\Omega_{\Lambda}$= 0.7, and 
$H_{\rm 0} = 70 {\rm \,km\, s^{-1}\, Mpc^{-1}}$. 
We summarize the major conclusions of this study as follows: \\

(1)  In our survey we have found 409 $IA$-excess objects 
whose rest-frame Ly$\alpha$ emission-line equivalent widths are greater than 20 \AA.
Applying the photometric redshift technique, we obtain a sample of 
198 Ly$\alpha$ emitter candidates.
 
(2) From this LAE sample, we find that the number density of LAEs 
at $z \sim$ 3.3 -- 4.8 is $n$(LAE) $\sim 10^{-4}$ Mpc$^{-3}$
on average. 

(3) In order to investigate host galaxy properties of our LAEs,
we investigate  the rest-frame UV luminosity distributions
of our LAE sample. The LAEs found in our study
are fainter than those found
for LAEs at $z \simeq$ 5.7 (Hu et al. 2004). 
We also find 
that the number density of LBGs is systematically
higher by a factor of $\approx$ 5 than LAEs in any UV luminosity.

(4) Using the observed Ly$\alpha$ luminosity,
we estimate the SFR of our LAEs ranging from 1.8 to 17.0 
$M_{\odot}$ yr$^{-1}$ with 
a median of 4.1 $M_{\odot}$ yr$^{-1}$.
We also estimate the SFR using the rest-frame UV luminosity.
Comparing these two SFRs, we find that the average
$SFR$(Ly$\alpha$)/$SFR$(UV) ratios  range from 0.5 to 2.
This ratio is nearly constant between $z \sim 3$ and 5. 

(5) Finally, we investigate the cosmic star formation history
based on the star formation rate density (SFRD).
We obtain $SFRD \sim 10^{-3.6} ~ M_{\odot}$ yr$^{-1}\, {\rm  Mpc}^{-3}$ 
on average. Although we find a systematic decrease of $SFRD$ with
increasing redshift for all LAEs found in our survey, the
SFRDs derived by using only bright LAEs
(log $L$(Ly$\alpha$) $> 42.6$ appear to be nearly constant between $z=3.3$ and 4.8. 

We would like to thank  the Subaru Telescope staff and the SXDS team for
their invaluable assistance. We would also like to thank 
K. Kodaira, S. Okamura, T. Yamada, K. Ohta, K. Shimasaku,
and M. Ouchi for many useful suggestions, comments, and 
encouragement during the course of this study.
We also thank T. Hayashino and H. Tamura for their kind technical
help for construction of the IA filter system.
This work was financially supported in part by
the Ministry of Education, Culture, Sports, Science and Technology
(Nos. 10044052, and 10304013) and JSPS (No. 15340059).
MA, SSS, and TN are JSPS fellows.
IRAF is distributed by the National Optical Astronomy Observatories,
which are operated by the Association of Universities for Research 
in Astronomy, Inc., under cooperative agreement with the National 
Science Foundation.

\newpage

\appendix
\section{Accuracy of our selection of LAEs using photometric redshifts}

Since we select LAEs from IA-excess objects applying the photometric redshift method, 
it is important to examine if there is no significant selection effect as a function of 
redshift. In order to demonstrate it, we perform computer simulations 
in the following way. 
By comparing differences between the true (model) and estimeated (photometric) 
redshifts, we can estimate the accuracy of our classification. 

First, we make the simulated data from the galaxy SEDs taking account of the photometric errors. 
The galaxy SEDs are generated by using the population synthesis model, GALAXEV 
(Bruzual \& Charlot 2003). 
To make various kinds of SEDs with strong emission line, we calculate the emission-line 
fluxes as a function of $EW_0({\rm H}\alpha)$, $EW_0$([OIII]), $EW_0$([OII]), and $EW_0({\rm Ly} \alpha)$ 
instead of being proportional to the ionizing photon production rate. 
We adopt three cases of $EW_0$ for each line: $EW_0=100$, 200, and 400 \AA~
for low-{\it z} emission line galaxies, $EW_0=65$, 130, and 260 \AA~ for LAEs. 
Second, we select emission-line galaxies from simulated catalogs using the same 
criteria written in section 3.2. 
Third, we apply the photometric redshift method to the simulated data and 
compare the photometric redshift ($z_{\rm phot}$) with the model redshifts ($z_{\rm model}$). 

Comparisons between $z_{\rm model}$ and $z_{\rm phot}$ are shown in figure 15. 
All the model LAEs are selected as LAEs using our photometric redshift method. 
Although some of the model low-{\it z} emission line galaxies are selected as LAEs, 
the fraction of misclassification is less than 10 \% for the all IA-filter bands. 
We, therefore, conclude that there is no significant selection effect as a 
function of redshift.


\clearpage
\begin{longtable}{rccrccc}
  \caption{The journal of observation}\label{tab:first}
\hline
\hline
Band & $\lambda_{\rm c}$ & $FWHM$ & & Obs. Date & Total Integ. Time & $M_{\rm lim}$(AB)\\
 & (\AA~) & (\AA~) & & & (s) & (3$\sigma$, $2^{\prime\prime}\phi$) \\
\endfirsthead
\hline
\endhead
\hline
$B$ & 4364 &  1008	&& 2000 Nov 24,25	& 	&  \\
			 &		& 		&& 2002 Jan 13		&	&  \\
	Total	 & 		& 		&& 					& 18000 & 28.2 \\
\hline
$R_c$ & 6410	& 1576	&& 2000 Aug 1		& 	&  \\
			 & 		& 		&& 2000 Nov 21-24	& 	&  \\
			 &		&		&& 2001 Nov 17		& 	&  \\
			 &		& 		&& 2002 Jan 13		&	&  \\
	Total	 &		& 		&& 					& 12000	& 27.4 \\
\hline
$i^{\prime}$ & 7589	& 1535	&& 2000 Nov 25		& 	& \\
			 		 & 		& 		&& 2002 Jan 13		&	& \\
	Total	 		 & 		& 		&& 					& 13200	& 27.0 \\
\hline
$z^{\prime}$ & 9024 & 1409	&& 2001 Oct 15,19,20&  5700	& 25.8 \\
\hline
$IA527$		 & 5272 & 242	&& 2002 Nov 6		& 5280 	& 26.8 \\
$IA574$		 & 5743 & 271	&& 2002 Nov 6		& 7200	& 26.5 \\
$IA598$		 & 6000	& 294	&& 2002 Nov 6		& 6720	& 26.5 \\
$IA624$		 & 6226 & 299	&& 2002 Oct 30,31	& 10560	& 26.7 \\
$IA651$		 & 6502	& 322	&& 2002 Oct 30,31	& 9600	& 26.7 \\
$IA679$		 & 6788	& 336	&& 2002 Oct 30,31	& 10560	& 26.8 \\
$IA709$		 & 7082	& 318	&& 2002 Oct 30,31	& 11520	& 26.6 \\
\hline
\end{longtable}

\clearpage

\begin{table}
  \caption{The numbers of  LAE ($EW_{\rm rest} > 20$ \AA) candidates.}\label{tab:second}
  \begin{center}
    \begin{tabular}{lccccc}
\hline
\hline
Filter & $z({\rm Ly \alpha})$ & $EW_{\rm obs}$ & $D[IA]-IA$ & IA-excess objects\footnotemark[$\dagger$]  & LAE candidates\footnotemark[$\ddagger$]  \\
       &                      & ($EW_{\rm rest}=20$\AA) & (mag) & & \\
\hline
IA527 & 3.336&  86.711 &  0.332 & 74 & 64 \\
IA574 & 3.723&  94.457 &  0.325 & 47 & 40 \\
IA598 & 3.934&  98.684 &  0.314 & 42 & 20 \\
IA624 & 4.120& 102.401 &  0.320 & 48 & 26 \\
IA651 & 4.347& 106.941 &  0.311 & 60 & 16 \\
IA679 & 4.582& 111.645 &  0.311 & 77 & 17 \\
IA709 & 4.824& 116.480 &  0.339 & 61 & 15 \\
\hline
\multicolumn{6}{l}{\footnotemark[$\dagger$] {The number of IA-excess objects (see section 3.2).}}\\
\multicolumn{6}{l}{\footnotemark[$\ddagger$] {The number of LAEs (see section 3.3).}}\\
    \end{tabular}
  \end{center}
\end{table}

\clearpage


\begin{longtable}{r@{ }r@{ }r@{ }r@{ }r@{ }r@{ }r@{ }r@{ }r@{ }r@{ }r@{ }r@{ }r@{ }r}
  \caption{Catalog of LAE candidates}\label{tab:third}
  \hline\hline
   ID   & $\Delta$RA & $\Delta$DEC & $B$ & $R$ & $i^\prime$ & $z^\prime$ & IA527 & IA574 & IA598 & IA624 & IA651 & IA679 & IA709 \\
        & (arcmin)   & (arcmin)    &     &     &            &            &       &       &       &       &       &       &       \\
\hline
\endfirsthead
  \hline\hline
   ID   & $\Delta$RA & $\Delta$DEC & $B$ & $R$ & $i^\prime$ & $z^\prime$ & IA527 & IA574 & IA598 & IA624 & IA651 & IA679 & IA709 \\
        & (arcmin)   & (arcmin)    &     &     &            &            &       &       &       &       &       &       &       \\
\hline
\endhead
  \hline
\endfoot
  \hline
\endlastfoot
\hline
\multicolumn{14}{c}{IA527}\\
\hline
   7048 & 31.12 &  2.10 &$ 24.9$ &$ 24.5$ &$ 24.6$ &$ 24.7$ & 24.1 &$ 24.7$ &$ 24.5$ &$ 24.5$ &$ 24.6$ &$ 24.6$ &$ 24.6$\\
   7988 & 17.98 &  2.26 &$ 27.6$ &$ 26.8$ &$ 27.1$ &$>26.2$ & 25.5 &$>26.9$ &$>26.9$ &$ 27.0$ &$ 26.7$ &$ 27.0$ &$ 26.6$\\
   9442 &  4.69 &  2.65 &$ 26.0$ &$ 25.3$ &$ 25.3$ &$ 25.3$ & 25.0 &$ 25.4$ &$ 25.4$ &$ 25.5$ &$ 25.4$ &$ 25.4$ &$ 25.4$\\
   9620 &  3.87 &  2.62 &$ 27.5$ &$ 27.4$ &$>27.4$ &$>26.2$ & 26.2 &$>26.9$ &$>26.9$ &$>27.1$ &$ 27.0$ &$>27.2$ &$>27.0$\\
  11523 & 32.95 &  3.08 &$ 26.7$ &$ 26.2$ &$ 26.4$ &$>26.2$ & 25.4 &$ 26.3$ &$ 26.4$ &$ 26.0$ &$ 26.4$ &$ 26.2$ &$ 26.7$\\
  11538 & 29.70 &  3.08 &$ 27.2$ &$ 26.6$ &$ 26.5$ &$>26.2$ & 25.5 &$ 26.5$ &$>26.9$ &$ 27.0$ &$ 26.8$ &$>27.2$ &$ 26.8$\\
  13069 & 15.33 &  3.46 &$ 26.8$ &$ 26.1$ &$ 26.3$ &$>26.2$ & 25.5 &$ 26.2$ &$ 26.3$ &$ 26.1$ &$ 26.1$ &$ 26.3$ &$ 26.2$\\
  15427 & 14.95 &  3.99 &$ 28.4$ &$>27.8$ &$>27.4$ &$>26.2$ & 26.1 &$>26.9$ &$>26.9$ &$>27.1$ &$>27.1$ &$>27.2$ &$>27.0$\\
  15823 & 13.32 &  4.09 &$ 27.1$ &$ 26.7$ &$ 27.0$ &$>26.2$ & 25.8 &$ 26.2$ &$ 26.4$ &$ 27.0$ &$ 26.8$ &$ 26.7$ &$ 26.6$\\
  17533 &  7.25 &  4.47 &$ 27.1$ &$ 26.5$ &$ 26.3$ &$>26.2$ & 25.6 &$ 26.8$ &$ 26.5$ &$ 26.2$ &$ 26.9$ &$ 26.9$ &$ 26.7$\\
  19286 & 16.25 &  4.90 &$ 27.4$ &$ 26.8$ &$ 26.6$ &$>26.2$ & 25.7 &$ 26.7$ &$ 26.3$ &$>27.1$ &$ 27.0$ &$>27.2$ &$ 26.6$\\
  27691 &  3.68 &  6.87 &$>28.6$ &$>27.8$ &$>27.4$ &$>26.2$ & 25.8 &$>26.9$ &$>26.9$ &$>27.1$ &$>27.1$ &$>27.2$ &$>27.0$\\
  28584 & 32.53 &  7.12 &$ 26.1$ &$ 25.8$ &$ 25.9$ &$ 25.9$ & 25.3 &$ 25.8$ &$ 26.0$ &$ 25.8$ &$ 25.6$ &$ 26.0$ &$ 25.8$\\
  28996 & 32.98 &  7.22 &$ 26.1$ &$ 25.7$ &$ 25.7$ &$ 25.3$ & 24.9 &$ 25.8$ &$ 25.9$ &$ 25.6$ &$ 25.5$ &$ 25.8$ &$ 25.6$\\
  29083 & 17.27 &  7.19 &$ 27.7$ &$ 26.7$ &$ 26.6$ &$>26.2$ & 25.9 &$>26.9$ &$ 26.8$ &$ 26.6$ &$ 26.6$ &$ 26.9$ &$ 26.7$\\
  37812 & 13.88 &  9.34 &$ 26.8$ &$ 25.9$ &$ 25.8$ &$ 25.6$ & 25.3 &$ 26.5$ &$ 25.8$ &$ 25.6$ &$ 25.7$ &$ 26.0$ &$ 25.9$\\
  38785 & 23.54 &  9.45 &$ 26.6$ &$ 26.0$ &$ 26.0$ &$>26.2$ & 25.4 &$ 26.5$ &$ 26.1$ &$ 26.8$ &$ 26.3$ &$ 26.3$ &$ 26.1$\\
  41668 & 27.31 & 10.14 &$ 27.6$ &$ 26.8$ &$ 26.9$ &$>26.2$ & 25.8 &$>26.9$ &$ 26.8$ &$>27.1$ &$>27.1$ &$ 27.0$ &$>27.0$\\
  44636 & 10.55 & 10.97 &$ 26.3$ &$ 25.9$ &$ 25.7$ &$ 26.0$ & 25.4 &$ 25.7$ &$ 26.4$ &$ 26.0$ &$ 26.4$ &$ 25.9$ &$ 26.0$\\
  47969 & 28.10 & 11.70 &$ 26.3$ &$ 25.5$ &$ 25.4$ &$ 25.3$ & 24.3 &$ 25.5$ &$ 25.5$ &$ 25.9$ &$ 25.3$ &$ 25.6$ &$ 25.5$\\
  49443 &  5.92 & 12.04 &$>28.6$ &$ 27.1$ &$ 27.1$ &$>26.2$ & 26.0 &$>26.9$ &$>26.9$ &$>27.1$ &$ 26.5$ &$>27.2$ &$>27.0$\\
  51844 & 27.84 & 12.52 &$ 27.6$ &$ 26.9$ &$ 27.1$ &$>26.2$ & 25.6 &$>26.9$ &$>26.9$ &$>27.1$ &$>27.1$ &$>27.2$ &$ 26.6$\\
  54027 &  8.21 & 13.04 &$ 27.0$ &$ 25.8$ &$ 25.6$ &$ 25.5$ & 25.2 &$ 26.1$ &$ 25.7$ &$ 25.7$ &$ 25.7$ &$ 25.6$ &$ 25.7$\\
  55020 & 22.84 & 13.23 &$ 27.9$ &$ 27.3$ &$ 27.4$ &$>26.2$ & 26.0 &$>26.9$ &$>26.9$ &$>27.1$ &$>27.1$ &$>27.2$ &$>27.0$\\
  55100 & 25.17 & 13.25 &$ 26.9$ &$ 26.7$ &$ 26.5$ &$ 26.1$ & 25.8 &$ 26.3$ &$>26.9$ &$ 26.9$ &$>27.1$ &$>27.2$ &$ 26.9$\\
  55848 & 20.90 & 13.43 &$ 27.1$ &$ 26.4$ &$ 26.3$ &$>26.2$ & 24.7 &$ 26.1$ &$>26.9$ &$ 26.9$ &$ 26.4$ &$ 26.6$ &$ 26.9$\\
  57223 & 13.32 & 13.75 &$ 28.5$ &$ 27.4$ &$>27.4$ &$>26.2$ & 26.0 &$>26.9$ &$>26.9$ &$>27.1$ &$>27.1$ &$>27.2$ &$>27.0$\\
  57844 & 30.39 & 13.89 &$ 27.0$ &$ 26.0$ &$ 26.2$ &$>26.2$ & 25.4 &$>26.9$ &$ 26.6$ &$ 26.3$ &$ 26.1$ &$ 26.3$ &$ 26.7$\\
  59074 & 24.21 & 14.14 &$>28.6$ &$ 27.6$ &$>27.4$ &$>26.2$ & 26.2 &$>26.9$ &$>26.9$ &$>27.1$ &$>27.1$ &$>27.2$ &$>27.0$\\
  59405 & 24.81 & 14.24 &$ 25.7$ &$ 25.4$ &$ 25.4$ &$ 25.7$ & 24.7 &$ 25.4$ &$ 25.3$ &$ 25.5$ &$ 25.3$ &$ 25.5$ &$ 25.4$\\
  60224 & 12.82 & 14.45 &$ 27.1$ &$ 26.4$ &$ 26.5$ &$>26.2$ & 25.5 &$ 26.4$ &$>26.9$ &$ 26.2$ &$ 26.6$ &$ 26.5$ &$ 26.5$\\
  60455 & 16.47 & 14.52 &$ 26.4$ &$ 25.7$ &$ 25.6$ &$ 25.9$ & 25.2 &$ 25.7$ &$ 25.9$ &$ 25.6$ &$ 25.6$ &$ 25.8$ &$ 25.8$\\
  60664 & 12.93 & 14.52 &$ 27.4$ &$ 26.7$ &$ 26.9$ &$>26.2$ & 25.9 &$ 26.8$ &$>26.9$ &$ 26.9$ &$ 26.1$ &$ 27.1$ &$>27.0$\\
  63293 & 13.47 & 15.21 &$ 27.3$ &$ 26.5$ &$ 26.4$ &$>26.2$ & 25.8 &$ 26.8$ &$ 26.4$ &$ 26.4$ &$ 26.5$ &$ 26.3$ &$ 26.5$\\
  69085 & 10.84 & 16.51 &$ 27.5$ &$ 27.4$ &$ 27.0$ &$>26.2$ & 26.2 &$>26.9$ &$ 26.7$ &$>27.1$ &$ 26.7$ &$ 26.8$ &$>27.0$\\
  70551 & 27.75 & 16.89 &$ 26.7$ &$ 26.0$ &$ 26.0$ &$ 25.7$ & 25.2 &$ 25.9$ &$ 25.9$ &$ 26.1$ &$ 25.8$ &$ 25.9$ &$ 26.0$\\
  71679 & 23.43 & 17.15 &$ 26.7$ &$ 25.9$ &$ 25.9$ &$ 25.7$ & 25.0 &$ 26.2$ &$ 26.0$ &$ 25.6$ &$ 25.9$ &$ 25.8$ &$ 25.9$\\
  72094 & 15.20 & 17.23 &$ 26.5$ &$ 25.8$ &$ 25.8$ &$ 25.8$ & 25.4 &$ 26.1$ &$ 25.9$ &$ 25.8$ &$ 26.0$ &$ 25.9$ &$ 25.8$\\
  72647 &  9.68 & 17.36 &$ 27.4$ &$ 26.9$ &$ 27.1$ &$>26.2$ & 25.6 &$ 26.4$ &$>26.9$ &$ 26.7$ &$ 26.6$ &$>27.2$ &$ 26.7$\\
  74055 & 19.68 & 17.67 &$ 27.8$ &$ 27.4$ &$ 27.3$ &$>26.2$ & 25.9 &$>26.9$ &$>26.9$ &$>27.1$ &$>27.1$ &$ 27.0$ &$>27.0$\\
  74299 &  9.17 & 17.72 &$ 27.6$ &$ 27.6$ &$ 26.5$ &$>26.2$ & 26.1 &$>26.9$ &$>26.9$ &$ 26.6$ &$>27.1$ &$>27.2$ &$>27.0$\\
  75353 & 18.23 & 17.99 &$ 27.6$ &$ 26.9$ &$ 27.1$ &$>26.2$ & 25.7 &$>26.9$ &$ 26.8$ &$>27.1$ &$ 26.9$ &$ 27.2$ &$>27.0$\\
  77460 & 23.33 & 18.48 &$ 27.0$ &$ 26.1$ &$ 26.0$ &$ 26.0$ & 25.4 &$ 25.9$ &$ 26.4$ &$ 26.0$ &$ 26.1$ &$ 26.1$ &$ 26.1$\\
  78162 &  9.64 & 18.67 &$ 25.6$ &$ 24.8$ &$ 24.8$ &$ 25.1$ & 24.3 &$ 24.9$ &$ 24.9$ &$ 24.8$ &$ 24.8$ &$ 24.8$ &$ 24.8$\\
  80388 & 13.88 & 19.15 &$ 27.9$ &$ 27.4$ &$>27.4$ &$ 26.2$ & 26.0 &$>26.9$ &$>26.9$ &$ 26.6$ &$>27.1$ &$>27.2$ &$ 26.8$\\
  81380 & 33.07 & 19.44 &$ 26.8$ &$ 25.8$ &$ 25.6$ &$ 25.9$ & 25.3 &$ 26.0$ &$ 25.9$ &$ 25.8$ &$ 25.6$ &$ 25.8$ &$ 25.6$\\
  82145 &  2.78 & 19.63 &$ 27.6$ &$ 26.6$ &$ 27.1$ &$>26.2$ & 25.8 &$ 26.7$ &$ 26.8$ &$ 26.6$ &$>27.1$ &$>27.2$ &$>27.0$\\
  82756 & 22.50 & 19.73 &$ 27.6$ &$ 27.0$ &$ 27.3$ &$>26.2$ & 25.5 &$>26.9$ &$>26.9$ &$>27.1$ &$>27.1$ &$ 27.0$ &$ 26.6$\\
  83157 & 25.82 & 19.86 &$ 26.5$ &$ 25.5$ &$ 25.6$ &$ 25.8$ & 24.5 &$ 25.9$ &$ 26.0$ &$ 25.7$ &$ 25.4$ &$ 25.6$ &$ 25.4$\\
  86220 & 11.33 & 20.53 &$ 27.2$ &$ 26.5$ &$ 26.5$ &$>26.2$ & 25.8 &$ 26.4$ &$ 26.7$ &$ 26.4$ &$ 26.7$ &$>27.2$ &$>27.0$\\
  87141 &  4.41 & 20.77 &$ 26.7$ &$ 25.8$ &$ 25.6$ &$ 25.4$ & 24.9 &$ 26.0$ &$ 25.9$ &$ 26.1$ &$ 25.7$ &$ 26.0$ &$ 25.8$\\
  87553 & 11.80 & 20.85 &$ 27.1$ &$ 26.4$ &$ 26.7$ &$>26.2$ & 25.2 &$>26.9$ &$>26.9$ &$ 26.1$ &$ 26.5$ &$ 26.2$ &$ 26.7$\\
  88785 & 30.56 & 21.13 &$ 27.2$ &$ 26.1$ &$ 26.1$ &$ 26.1$ & 25.5 &$ 26.0$ &$ 26.1$ &$ 26.2$ &$ 26.1$ &$ 26.4$ &$ 26.1$\\
  89485 & 24.51 & 21.30 &$ 27.0$ &$ 26.2$ &$ 26.2$ &$>26.2$ & 25.5 &$ 26.3$ &$ 26.1$ &$ 26.1$ &$ 26.2$ &$ 26.4$ &$ 26.0$\\
  89896 &  2.73 & 21.38 &$>28.6$ &$>27.8$ &$>27.4$ &$>26.2$ & 26.1 &$>26.9$ &$>26.9$ &$>27.1$ &$ 27.0$ &$>27.2$ &$>27.0$\\
  90874 & 28.10 & 21.69 &$ 26.8$ &$ 26.4$ &$ 26.3$ &$>26.2$ & 25.4 &$>26.9$ &$ 26.3$ &$ 26.6$ &$ 26.2$ &$ 26.6$ &$ 26.3$\\
  94639 & 10.37 & 22.51 &$ 28.0$ &$ 27.4$ &$ 27.2$ &$>26.2$ & 25.8 &$>26.9$ &$>26.9$ &$>27.1$ &$>27.1$ &$>27.2$ &$>27.0$\\
  94941 &  9.14 & 22.58 &$ 28.4$ &$>27.8$ &$>27.4$ &$>26.2$ & 26.1 &$>26.9$ &$>26.9$ &$>27.1$ &$>27.1$ &$>27.2$ &$>27.0$\\
  95272 & 30.36 & 22.69 &$>28.6$ &$ 27.6$ &$>27.4$ &$>26.2$ & 26.1 &$>26.9$ &$>26.9$ &$>27.1$ &$>27.1$ &$>27.2$ &$>27.0$\\
  96625 & 12.26 & 22.98 &$ 26.5$ &$ 26.2$ &$ 26.2$ &$>26.2$ & 25.5 &$>26.6$ &$ 26.3$ &$ 26.1$ &$ 26.4$ &$ 26.4$ &$ 26.1$\\
  99496 & 33.17 & 23.64 &$ 27.4$ &$ 27.2$ &$>27.4$ &$>26.2$ & 25.7 &$>26.9$ &$>26.9$ &$ 26.7$ &$>27.1$ &$>27.2$ &$ 26.7$\\
 100145 & 24.45 & 23.81 &$ 27.5$ &$ 26.7$ &$ 26.8$ &$>26.2$ & 25.6 &$>26.9$ &$ 26.7$ &$>27.1$ &$ 26.4$ &$ 27.1$ &$ 26.6$\\
 101147 &  3.67 & 24.04 &$ 27.8$ &$ 27.0$ &$ 26.4$ &$ 25.8$ & 26.0 &$>26.9$ &$>26.9$ &$>27.1$ &$>27.1$ &$ 26.9$ &$>27.0$\\
 102766 & 10.22 & 24.38 &$ 27.2$ &$ 26.7$ &$ 26.7$ &$>26.2$ & 25.7 &$>26.9$ &$ 26.4$ &$>27.1$ &$>27.1$ &$>27.2$ &$ 26.7$\\
\hline
\multicolumn{14}{c}{IA574}\\
\hline                   
   6580 & 23.34 &  2.02 &$ 26.8$ &$ 25.3$ &$ 25.4$ &$ 25.1$ &$ 26.1$ & 24.9 &$ 26.2$ &$ 25.4$ &$ 25.5$ &$ 25.5$ &$ 25.4$\\
  10969 &  9.65 &  3.06 &$ 27.0$ &$ 25.4$ &$ 25.5$ &$ 25.8$ &$ 26.0$ & 24.5 &$ 25.4$ &$ 25.6$ &$ 25.6$ &$ 25.4$ &$ 25.4$\\
  11379 &  3.01 &  3.15 &$ 28.5$ &$ 26.7$ &$ 26.4$ &$>26.2$ &$>27.2$ & 25.7 &$ 26.7$ &$>27.1$ &$ 26.9$ &$ 27.1$ &$>27.0$\\
  13034 & 31.65 &  3.51 &$ 28.4$ &$ 27.5$ &$>27.4$ &$>26.2$ &$>27.2$ & 26.0 &$>26.9$ &$ 26.9$ &$>27.1$ &$>27.2$ &$>27.0$\\
  21251 & 14.75 &  5.62 &$>28.6$ &$ 27.5$ &$ 27.3$ &$>26.2$ &$>27.2$ & 25.9 &$>26.9$ &$>27.1$ &$>27.1$ &$>27.2$ &$ 26.9$\\
  22223 & 11.00 &  5.85 &$ 27.5$ &$ 26.5$ &$ 26.4$ &$>26.2$ &$ 26.7$ & 25.6 &$>26.9$ &$ 26.3$ &$ 26.5$ &$ 26.8$ &$ 26.3$\\
  25092 & 16.05 &  6.57 &$ 27.9$ &$ 26.5$ &$ 26.7$ &$>26.2$ &$>27.2$ & 25.7 &$ 26.4$ &$ 26.5$ &$ 26.8$ &$ 26.8$ &$ 26.5$\\
  34695 & 19.17 &  9.03 &$ 27.2$ &$ 25.7$ &$ 25.5$ &$ 25.8$ &$ 26.6$ & 24.7 &$ 25.7$ &$ 25.8$ &$ 25.7$ &$ 25.7$ &$ 25.7$\\
  36987 & 15.02 &  9.52 &$>28.6$ &$ 27.3$ &$ 27.1$ &$>26.2$ &$>27.2$ & 25.8 &$ 26.6$ &$ 26.4$ &$>27.1$ &$ 27.1$ &$>27.0$\\
  45242 & 28.40 & 11.64 &$ 27.9$ &$ 26.6$ &$ 26.5$ &$>26.2$ &$ 26.7$ & 25.7 &$>26.9$ &$ 26.8$ &$ 26.5$ &$ 26.7$ &$>27.0$\\
  46603 & 32.09 & 11.96 &$ 27.7$ &$ 26.3$ &$ 25.7$ &$ 25.5$ &$>27.2$ & 24.4 &$ 26.7$ &$ 26.4$ &$ 26.4$ &$ 26.0$ &$ 25.7$\\
  51420 & 28.60 & 13.02 &$ 27.6$ &$ 25.9$ &$ 26.0$ &$ 25.5$ &$ 26.7$ & 25.2 &$ 25.8$ &$ 25.9$ &$ 25.8$ &$ 26.1$ &$ 25.9$\\
  53464 & 29.48 & 13.53 &$ 27.2$ &$ 25.7$ &$ 25.8$ &$ 26.1$ &$ 26.6$ & 25.0 &$ 25.6$ &$ 25.8$ &$ 25.7$ &$ 25.7$ &$ 25.7$\\
  53930 &  3.36 & 13.66 &$ 27.5$ &$ 25.5$ &$ 25.6$ &$>26.2$ &$ 26.4$ & 24.8 &$ 25.1$ &$ 25.6$ &$ 25.5$ &$ 25.8$ &$ 25.8$\\
  54185 & 18.45 & 13.68 &$ 28.2$ &$ 26.4$ &$ 26.4$ &$>26.2$ &$>27.2$ & 25.6 &$ 26.0$ &$ 26.5$ &$ 26.4$ &$ 26.6$ &$ 26.3$\\
  55740 &  5.22 & 14.04 &$ 28.5$ &$ 26.4$ &$ 26.3$ &$ 26.0$ &$>27.2$ & 25.6 &$ 25.4$ &$ 26.7$ &$ 26.5$ &$ 26.8$ &$ 26.7$\\
  55786 & 21.14 & 14.03 &$ 27.6$ &$ 26.3$ &$ 26.1$ &$>26.2$ &$ 26.6$ & 25.5 &$ 26.0$ &$>27.1$ &$ 26.9$ &$ 26.3$ &$ 26.5$\\
  57174 & 11.44 & 14.36 &$>28.6$ &$ 26.9$ &$ 27.0$ &$>26.2$ &$>27.2$ & 25.8 &$>26.9$ &$>27.1$ &$ 26.8$ &$>27.2$ &$>27.0$\\
  58770 & 31.72 & 14.78 &$ 27.7$ &$ 26.1$ &$ 25.9$ &$>26.2$ &$ 26.3$ & 25.3 &$ 26.3$ &$ 25.9$ &$ 26.4$ &$ 26.3$ &$ 26.2$\\
  65463 & 27.52 & 16.41 &$ 27.1$ &$ 25.8$ &$ 25.7$ &$ 25.6$ &$ 26.9$ & 25.2 &$ 25.8$ &$ 25.7$ &$ 25.7$ &$ 25.6$ &$ 25.8$\\
  65839 & 15.68 & 16.47 &$ 27.5$ &$ 25.7$ &$ 25.9$ &$ 25.8$ &$ 26.8$ & 25.2 &$ 25.0$ &$ 25.7$ &$ 25.9$ &$ 25.9$ &$ 25.8$\\
  66933 & 17.41 & 16.71 &$ 27.8$ &$ 26.7$ &$ 26.5$ &$ 26.1$ &$ 27.1$ & 25.8 &$ 26.1$ &$ 26.5$ &$ 26.5$ &$ 26.5$ &$ 26.8$\\
  68446 & 22.09 & 17.10 &$ 28.1$ &$ 26.2$ &$ 26.3$ &$>26.2$ &$>27.2$ & 25.5 &$ 26.5$ &$ 26.4$ &$ 26.1$ &$ 26.3$ &$ 26.3$\\
  69065 & 26.74 & 17.21 &$>28.6$ &$ 27.8$ &$ 27.3$ &$>26.2$ &$>27.2$ & 26.0 &$>26.9$ &$ 26.8$ &$ 27.0$ &$>27.2$ &$>27.0$\\
  71041 & 20.31 & 17.71 &$ 26.8$ &$ 25.4$ &$ 25.2$ &$ 25.2$ &$ 26.2$ & 24.7 &$ 25.4$ &$ 25.5$ &$ 25.3$ &$ 25.3$ &$ 25.4$\\
  72707 & 26.20 & 18.09 &$ 28.4$ &$ 27.7$ &$>27.4$ &$>26.2$ &$>27.2$ & 26.0 &$>26.9$ &$>27.1$ &$>27.1$ &$>27.2$ &$>27.0$\\
  73613 &  8.82 & 18.37 &$ 26.9$ &$ 25.4$ &$ 25.5$ &$ 25.7$ &$ 26.4$ & 24.5 &$ 25.5$ &$ 25.4$ &$ 25.4$ &$ 25.5$ &$ 25.6$\\
  74326 &  8.84 & 18.54 &$ 27.1$ &$ 25.6$ &$ 25.6$ &$ 26.1$ &$ 26.2$ & 25.2 &$ 25.8$ &$ 25.8$ &$ 25.7$ &$ 25.9$ &$ 25.6$\\
  75127 & 21.00 & 18.69 &$ 26.7$ &$ 25.5$ &$ 25.5$ &$ 25.9$ &$ 26.5$ & 24.9 &$ 25.5$ &$ 25.5$ &$ 25.5$ &$ 25.7$ &$ 25.5$\\
  75930 & 25.58 & 18.89 &$ 28.2$ &$ 27.1$ &$ 27.1$ &$>26.2$ &$>27.2$ & 25.5 &$>26.9$ &$>27.1$ &$ 26.5$ &$ 26.9$ &$>27.0$\\
  76201 & 11.48 & 18.96 &$ 27.3$ &$ 25.9$ &$ 25.9$ &$>26.2$ &$ 26.2$ & 25.0 &$ 26.0$ &$ 25.8$ &$ 25.9$ &$ 26.1$ &$ 25.8$\\
  77116 & 28.22 & 19.32 &$ 27.5$ &$ 26.2$ &$ 26.2$ &$>26.2$ &$>27.2$ & 25.4 &$ 26.4$ &$ 26.2$ &$ 26.0$ &$ 26.3$ &$ 26.7$\\
  78959 & 28.68 & 19.65 &$ 28.3$ &$ 27.0$ &$ 26.8$ &$>26.2$ &$>27.2$ & 25.2 &$>26.9$ &$ 27.0$ &$ 26.8$ &$>27.2$ &$ 26.8$\\
  79754 &  3.03 & 19.82 &$>28.6$ &$>27.8$ &$>27.4$ &$>26.2$ &$>27.2$ & 25.9 &$>26.9$ &$>27.1$ &$>27.1$ &$>27.2$ &$>27.0$\\
  82975 &  4.29 & 20.60 &$ 28.5$ &$>27.8$ &$>27.4$ &$>26.2$ &$>27.2$ & 26.0 &$>26.9$ &$ 26.8$ &$ 26.6$ &$>27.2$ &$>27.0$\\
  85436 & 24.16 & 21.39 &$ 28.1$ &$ 27.4$ &$ 27.0$ &$>26.2$ &$>27.2$ & 25.9 &$>26.9$ &$>27.1$ &$>27.1$ &$ 27.1$ &$>27.0$\\
  90651 & 22.42 & 22.51 &$>28.6$ &$>27.8$ &$>27.4$ &$>26.2$ &$ 27.0$ & 25.8 &$>26.9$ &$>27.1$ &$>27.1$ &$>27.2$ &$ 26.7$\\
  91865 &  9.85 & 22.81 &$ 27.4$ &$ 25.7$ &$ 25.5$ &$ 25.3$ &$ 27.1$ & 25.0 &$ 25.7$ &$ 25.7$ &$ 25.8$ &$ 25.8$ &$ 25.5$\\
  96523 & 32.25 & 23.86 &$>28.6$ &$ 26.6$ &$ 26.7$ &$>26.2$ &$>27.2$ & 25.6 &$ 26.2$ &$>27.1$ &$ 26.3$ &$ 27.0$ &$>27.0$\\
  99338 & 14.02 & 24.48 &$ 28.4$ &$ 27.1$ &$ 27.2$ &$>26.2$ &$>27.2$ & 25.8 &$>26.9$ &$>27.1$ &$ 26.5$ &$ 27.0$ &$>27.0$\\
\hline
\multicolumn{14}{c}{IA598}\\
\hline
  10519 &  8.44 &  2.93 &$ 27.4$ & 25.9 &$ 26.0$ &$>26.2$ &$ 26.4$ &$ 26.0$ & 25.2 &$ 26.2$ &$ 25.9$ &$ 25.9$ &$ 26.5$\\
  12390 & 15.71 &  3.39 &$ 27.6$ & 25.3 &$ 25.5$ &$ 25.4$ &$ 26.7$ &$ 26.3$ & 24.9 &$ 25.3$ &$ 25.5$ &$ 25.4$ &$ 25.5$\\
  13299 &  7.22 &  3.65 &$ 28.0$ & 25.7 &$ 26.0$ &$>26.2$ &$ 26.5$ &$ 25.9$ & 25.1 &$ 25.9$ &$ 25.9$ &$ 26.0$ &$ 25.8$\\
  27435 & 21.87 &  7.15 &$ 28.3$ & 26.6 &$ 26.4$ &$>26.2$ &$>27.2$ &$ 26.5$ & 25.4 &$>27.1$ &$ 26.7$ &$ 26.7$ &$ 26.8$\\
  33977 & 23.33 &  8.70 &$ 28.2$ & 26.3 &$ 26.4$ &$ 25.7$ &$>27.2$ &$ 25.7$ & 25.6 &$ 26.4$ &$ 26.5$ &$>27.2$ &$ 26.2$\\
  37912 & 16.71 &  9.77 &$ 27.5$ & 25.6 &$ 25.7$ &$ 25.6$ &$ 26.4$ &$ 25.7$ & 25.1 &$ 25.8$ &$ 25.7$ &$ 25.6$ &$ 25.8$\\
  53728 & 33.03 & 13.66 &$ 28.2$ & 26.3 &$ 26.4$ &$ 25.8$ &$ 27.1$ &$>26.9$ & 25.5 &$ 26.9$ &$ 26.8$ &$ 26.3$ &$ 26.3$\\
  53761 &  3.36 & 13.66 &$ 27.5$ & 25.5 &$ 25.6$ &$>26.2$ &$ 26.4$ &$ 24.8$ & 25.0 &$ 25.6$ &$ 25.6$ &$ 25.8$ &$ 25.8$\\
  55275 &  5.22 & 14.04 &$ 28.4$ & 26.3 &$ 26.3$ &$ 26.0$ &$>27.2$ &$ 25.6$ & 25.4 &$ 26.6$ &$ 26.4$ &$ 26.8$ &$ 26.7$\\
  58192 & 18.42 & 14.76 &$ 28.0$ & 25.6 &$ 25.7$ &$ 25.9$ &$ 26.1$ &$ 26.2$ & 25.1 &$ 25.7$ &$ 25.7$ &$ 26.0$ &$ 26.1$\\
  59826 & 15.24 & 15.13 &$ 28.4$ & 27.7 &$>27.4$ &$>26.2$ &$>27.2$ &$>26.9$ & 26.0 &$>27.1$ &$>27.1$ &$>27.2$ &$>27.0$\\
  60845 & 31.10 & 15.46 &$ 27.4$ & 25.5 &$ 25.7$ &$ 25.9$ &$ 26.5$ &$ 26.7$ & 25.1 &$ 25.0$ &$ 25.7$ &$ 25.9$ &$ 25.9$\\
  64675 & 15.68 & 16.46 &$ 27.6$ & 25.8 &$ 26.1$ &$ 25.9$ &$ 26.7$ &$ 25.3$ & 25.0 &$ 25.8$ &$ 25.9$ &$ 25.9$ &$ 25.9$\\
  65983 &  8.87 & 16.79 &$>28.6$ & 26.4 &$ 26.9$ &$>26.2$ &$ 27.1$ &$>26.9$ & 25.5 &$ 26.7$ &$ 26.6$ &$ 27.1$ &$ 26.8$\\
  80344 & 22.09 & 20.48 &$>28.6$ & 26.7 &$ 27.0$ &$>26.2$ &$>27.2$ &$>26.9$ & 25.5 &$ 26.9$ &$ 27.1$ &$ 26.8$ &$>27.0$\\
  81987 & 22.99 & 20.98 &$ 27.5$ & 26.2 &$ 26.3$ &$>26.2$ &$>27.2$ &$ 26.5$ & 25.5 &$ 26.4$ &$ 26.2$ &$ 26.6$ &$ 26.1$\\
  85081 & 27.30 & 21.70 &$ 27.6$ & 26.5 &$ 26.3$ &$ 25.8$ &$>27.2$ &$ 25.8$ & 25.6 &$>27.1$ &$ 26.4$ &$ 26.5$ &$ 26.3$\\
  87226 &  9.98 & 22.25 &$ 28.2$ & 26.0 &$ 26.1$ &$>26.2$ &$>27.2$ &$>26.9$ & 25.2 &$ 26.6$ &$ 26.4$ &$ 26.6$ &$ 26.1$\\
  89279 & 11.33 & 22.83 &$ 27.6$ & 24.7 &$ 24.3$ &$ 24.4$ &$ 26.3$ &$ 25.5$ & 24.1 &$ 25.1$ &$ 25.0$ &$ 24.8$ &$ 24.8$\\
  92776 & 32.35 & 23.61 &$ 28.5$ & 27.1 &$>27.4$ &$>26.2$ &$>27.2$ &$ 26.4$ & 25.9 &$>27.1$ &$>27.1$ &$>27.2$ &$ 26.7$\\
\hline                   
\multicolumn{14}{c}{IA651}\\
\hline
  18239 &  3.23 &  4.91 &$>28.6$ & 26.3 & 26.5 &$>26.2$ &$>27.2$ &$>26.9$ &$>26.9$ &$>27.1$ & 25.3 &$ 26.8$ &$ 26.4$\\
  18481 &  2.62 &  5.00 &$>28.6$ & 26.2 & 26.5 &$>26.2$ &$>27.2$ &$>26.9$ &$>26.9$ &$>27.1$ & 25.4 &$ 26.7$ &$ 27.0$\\
  18929 & 33.00 &  5.10 &$>28.6$ & 25.7 & 25.8 &$ 25.5$ &$>27.2$ &$>26.9$ &$>26.9$ &$>25.3$ & 25.3 &$ 25.9$ &$ 26.4$\\
  56248 & 15.97 & 14.13 &$>28.6$ & 25.5 & 25.7 &$ 25.7$ &$>27.2$ &$>26.9$ &$ 26.3$ &$ 25.5$ & 25.1 &$ 25.7$ &$ 25.8$\\
  58954 & 31.80 & 14.78 &$>28.6$ & 25.4 & 25.6 &$>26.2$ &$ 26.5$ &$ 26.2$ &$ 26.6$ &$ 25.8$ & 24.7 &$ 25.5$ &$ 25.4$\\
  60304 & 18.15 & 15.06 &$>28.6$ & 25.9 & 26.0 &$>26.2$ &$>27.2$ &$ 26.8$ &$ 26.6$ &$ 25.5$ & 25.4 &$ 25.9$ &$ 26.1$\\
  61944 & 27.74 & 15.45 &$ 28.3$ & 26.7 & 27.1 &$>26.2$ &$>27.2$ &$>26.9$ &$>26.9$ &$ 26.3$ & 25.9 &$ 27.2$ &$ 27.0$\\
  62565 & 19.97 & 15.59 &$>28.6$ & 26.1 & 26.2 &$>26.2$ &$>27.2$ &$ 26.8$ &$ 26.8$ &$ 26.3$ & 25.4 &$ 26.1$ &$ 26.2$\\
  65736 & 19.27 & 16.40 &$>28.6$ & 26.6 & 26.7 &$>26.2$ &$>27.2$ &$>26.9$ &$>26.9$ &$>27.1$ & 25.8 &$ 26.9$ &$ 26.5$\\
  68104 & 22.64 & 16.91 &$>28.6$ & 26.2 & 26.5 &$>26.2$ &$>27.2$ &$>26.9$ &$>26.9$ &$>27.1$ & 25.6 &$ 26.3$ &$ 26.6$\\
  70056 &  4.12 & 17.39 &$ 28.4$ & 26.0 & 26.1 &$ 25.5$ &$>27.2$ &$>26.9$ &$>26.9$ &$>27.1$ & 25.4 &$ 26.3$ &$ 26.7$\\
  87781 &  9.50 & 21.64 &$ 28.4$ & 25.7 & 25.7 &$ 25.7$ &$ 27.0$ &$ 26.6$ &$>26.9$ &$>27.1$ & 25.2 &$ 25.7$ &$ 25.8$\\
  95072 & 33.02 & 23.39 &$>28.6$ & 26.8 & 27.3 &$>26.2$ &$>27.2$ &$>26.9$ &$>26.9$ &$>27.1$ & 25.5 &$>27.2$ &$>27.0$\\
  96758 &  9.34 & 23.84 &$>28.6$ & 26.2 & 26.1 &$ 26.0$ &$ 26.6$ &$>26.9$ &$>26.9$ &$>27.1$ & 25.3 &$ 26.7$ &$ 26.0$\\
  97673 & 30.68 & 24.03 &$ 28.2$ & 26.1 & 26.2 &$>26.2$ &$>27.2$ &$>26.9$ &$>26.9$ &$>27.1$ & 25.4 &$ 25.7$ &$ 26.5$\\
  98943 & 14.27 & 24.33 &$>28.6$ & 25.6 & 25.6 &$>26.2$ &$>27.2$ &$ 26.5$ &$ 26.4$ &$ 26.7$ & 25.2 &$ 25.4$ &$ 25.7$\\
\hline
\multicolumn{14}{c}{IA679}\\
\hline
   9065 & 11.03 &  3.02 &$>28.6$ & 26.8 & 26.7 &$>26.2$ &$>27.2$ &$>26.9$ &$>26.9$ &$>27.1$ &$>27.1$ & 25.8 &$ 26.6$\\
  11843 & 12.88 &  3.78 &$>28.6$ & 25.1 & 24.8 &$ 24.6$ &$ 26.3$ &$ 25.6$ &$ 26.3$ &$ 26.0$ &$ 25.0$ & 24.1 &$ 25.1$\\
  16415 & 12.58 &  5.04 &$>28.6$ & 26.1 & 26.3 &$>26.2$ &$>27.2$ &$>26.9$ &$>26.9$ &$>27.1$ &$>27.1$ & 25.3 &$ 26.5$\\
  18878 & 19.37 &  5.73 &$>28.6$ & 26.9 & 26.9 &$>26.2$ &$>27.2$ &$>26.9$ &$>26.9$ &$>27.1$ &$ 26.7$ & 25.8 &$>27.0$\\
  19202 & 16.85 &  5.83 &$>28.6$ & 26.7 & 26.7 &$>26.2$ &$>27.2$ &$>26.9$ &$>26.9$ &$>27.1$ &$>27.1$ & 25.6 &$ 26.7$\\
  24769 & 18.54 &  7.37 &$>28.6$ & 25.5 & 25.6 &$ 25.5$ &$>27.2$ &$>26.9$ &$ 26.4$ &$>27.1$ &$ 26.9$ & 24.7 &$ 25.7$\\
  33768 &  2.80 &  9.83 &$>28.6$ & 26.1 & 25.9 &$ 26.1$ &$>27.2$ &$>26.9$ &$>26.9$ &$>27.1$ &$ 26.6$ & 25.4 &$ 25.8$\\
  46786 & 11.24 & 13.33 &$>28.6$ & 26.3 & 26.0 &$>26.2$ &$>27.2$ &$>26.9$ &$>26.9$ &$>27.1$ &$>27.1$ & 25.4 &$ 25.9$\\
  47730 & 19.04 & 13.55 &$>28.6$ & 26.2 & 25.8 &$ 25.7$ &$>27.2$ &$>26.9$ &$>26.9$ &$>26.5$ &$>27.1$ & 25.3 &$ 25.5$\\
  52102 & 18.29 & 14.62 &$>28.6$ & 26.2 & 26.3 &$ 26.1$ &$>27.2$ &$ 26.7$ &$ 26.8$ &$ 27.0$ &$ 26.0$ & 25.4 &$ 25.8$\\
  57794 & 18.81 & 16.05 &$ 28.4$ & 26.6 & 26.8 &$>26.2$ &$>27.2$ &$>26.9$ &$>26.9$ &$>27.1$ &$>27.1$ & 25.7 &$ 26.7$\\
  60173 & 24.90 & 16.68 &$>28.6$ & 25.9 & 25.6 &$ 25.7$ &$>27.2$ &$ 26.7$ &$>26.9$ &$>27.1$ &$ 26.8$ & 25.0 &$ 26.1$\\
  62116 & 18.75 & 17.17 &$>28.6$ & 26.1 & 25.9 &$ 26.1$ &$>27.2$ &$>26.9$ &$>26.6$ &$>27.1$ &$>27.1$ & 25.1 &$ 26.1$\\
  64479 & 13.46 & 17.75 &$>28.6$ & 26.3 & 26.2 &$ 25.7$ &$>27.2$ &$>26.9$ &$>26.6$ &$>26.3$ &$ 26.4$ & 25.5 &$ 26.0$\\
  65576 & 13.78 & 18.07 &$>28.6$ & 25.6 & 25.2 &$ 25.2$ &$>27.2$ &$ 26.6$ &$ 26.2$ &$ 26.3$ &$ 26.4$ & 24.8 &$ 25.2$\\
  74655 & 23.99 & 20.40 &$>28.6$ & 27.2 & 27.0 &$>26.2$ &$>27.2$ &$>26.9$ &$>26.9$ &$>27.1$ &$>27.1$ & 25.9 &$ 26.6$\\
  89098 & 32.65 & 24.33 &$>28.6$ & 26.2 & 26.4 &$ 26.2$ &$ 26.8$ &$>26.9$ &$>26.9$ &$>27.1$ &$>27.1$ & 25.4 &$ 26.2$\\
\hline
\multicolumn{14}{c}{IA709}\\
\hline
   8190 &  6.33 &  2.65 &$>28.6$ &$ 26.0$ & 25.5 &$ 26.0$ &$>27.2$ &$ 26.5$ &$>26.9$ &$>27.1$ &$ 26.2$ &$ 25.9$ & 25.1\\
  13683 &  3.29 &  4.01 &$>28.6$ &$>27.8$ & 27.3 &$>26.2$ &$>27.2$ &$>26.9$ &$>26.9$ &$>27.1$ &$>27.1$ &$>27.2$ & 25.8\\
  18701 & 13.61 &  5.25 &$>28.6$ &$ 27.2$ & 26.2 &$>26.2$ &$>27.2$ &$>26.9$ &$>26.9$ &$>27.1$ &$>27.1$ &$>27.2$ & 25.5\\
  31658 &  3.29 &  8.35 &$>28.6$ &$ 27.2$ & 26.0 &$>26.2$ &$>27.2$ &$>26.9$ &$>26.9$ &$>27.1$ &$>27.1$ &$>27.2$ & 25.3\\
  33815 & 21.05 &  8.91 &$>28.6$ &$ 26.2$ & 25.4 &$ 25.3$ &$>27.2$ &$>26.9$ &$>26.9$ &$ 26.8$ &$>27.1$ &$>27.2$ & 24.9\\
  34592 & 22.59 &  9.08 &$>28.6$ &$ 26.7$ & 26.0 &$>26.2$ &$>27.2$ &$>26.9$ &$>26.9$ &$>27.1$ &$>27.1$ &$>27.2$ & 25.3\\
  38261 & 20.65 &  9.95 &$>28.6$ &$ 26.8$ & 26.2 &$ 26.0$ &$>27.2$ &$>26.9$ &$>26.9$ &$ 27.0$ &$>27.1$ &$>27.2$ & 25.5\\
  51167 &  3.36 & 13.00 &$>28.6$ &$ 26.9$ & 25.8 &$ 25.8$ &$>27.2$ &$>26.9$ &$>26.6$ &$>27.1$ &$>27.1$ &$>27.2$ & 25.2\\
  55557 & 13.08 & 14.03 &$>28.6$ &$ 26.0$ & 25.5 &$ 26.1$ &$>27.2$ &$>26.9$ &$>26.9$ &$>27.1$ &$>27.1$ &$ 25.6$ & 25.0\\
  59617 &  4.65 & 14.98 &$>28.6$ &$ 25.9$ & 25.4 &$ 25.8$ &$>27.2$ &$>26.9$ &$>26.9$ &$>27.1$ &$>27.1$ &$>27.2$ & 24.4\\
  66451 &  9.95 & 16.61 &$>28.6$ &$ 26.5$ & 25.7 &$ 25.9$ &$>27.2$ &$>26.9$ &$>26.9$ &$>27.1$ &$>27.1$ &$>27.2$ & 25.2\\
  72647 & 25.39 & 18.09 &$>28.6$ &$ 27.0$ & 27.3 &$>26.2$ &$>27.2$ &$>26.9$ &$>26.9$ &$>27.1$ &$>27.1$ &$ 26.5$ & 25.8\\
  80850 & 29.77 & 20.09 &$>28.6$ &$ 26.6$ & 26.2 &$>26.2$ &$>27.2$ &$>26.9$ &$>26.9$ &$>27.1$ &$>27.1$ &$>27.2$ & 25.5\\
  83599 & 32.00 & 20.73 &$>28.6$ &$ 27.2$ & 27.2 &$>26.2$ &$>27.2$ &$>26.9$ &$>26.9$ &$>27.1$ &$>27.1$ &$>27.2$ & 25.7\\
  84928 & 11.99 & 21.07 &$>28.6$ &$ 26.8$ & 26.6 &$>26.2$ &$>27.2$ &$>26.9$ &$>26.9$ &$>27.1$ &$>27.1$ &$>27.2$ & 25.2\\
\end{longtable}


\clearpage

\begin{longtable}{rrrrrrrrr}
  \caption{Properties of LAE candidates}\label{tab:LTsample4}
  \hline\hline
   ID  & $EW_{\rm obs}$ & $EW_{\rm rest}$ & $f_{\rm Ly \alpha}$ & $L_{\rm Ly \alpha}$ & $L_\nu(UV)$ & $M_{UV}$ & $SFR_{\rm Ly \alpha}$ & $SFR_{UV}$ \\
       & (\AA)          & (\AA)           & (erg s$^{-1}$ cm$^{-2}$) & (erg s$^{-1}$) & (erg s$^{-1}$ Hz$^{-1}$) & (mag) & ($M_\odot {\rm yr}^{-1}$) & ($M_\odot {\rm yr}^{-1}$) \\
\endfirsthead
\hline \hline
   ID  & $EW_{\rm obs}$ & $EW_{\rm rest}$ & $f_{\rm Ly \alpha}$ & $L_{\rm Ly \alpha}$ & $L_\nu(UV)$ & $M_{UV}$ & $SFR_{\rm Ly \alpha}$ & $SFR_{UV}$ \\
       & (\AA)          & (\AA)           & (erg s$^{-1}$ cm$^{-2}$) & (erg s$^{-1}$) & (erg s$^{-1}$ Hz$^{-1}$) & (mag) & ($M_\odot {\rm yr}^{-1}$) & ($M_\odot {\rm yr}^{-1}$) \\
  \hline
\endhead
  \hline
\endfoot
  \hline
\endlastfoot
\hline
\multicolumn{9}{c}{IA527}\\
\hline
   7048&  112.7&   26.0&  6.97E-17&  6.95E+42&  1.32E+29& -21.2&   6.3&  18.4\\
   7988&  509.4&  117.5&  3.96E-17&  3.95E+42&  1.65E+28& -18.9&   3.6&   2.3\\
   9442&   89.5&   20.6&  2.56E-17&  2.56E+42&  6.10E+28& -20.4&   2.3&   8.5\\
   9620&  484.4&  111.7&  2.17E-17&  2.17E+42&  9.56E+27& -18.4&   2.0&   1.3\\
  11523&  284.2&   65.6&  3.62E-17&  3.61E+42&  2.71E+28& -19.5&   3.3&   3.8\\
  11538&  397.5&   91.7&  3.61E-17&  3.60E+42&  1.93E+28& -19.1&   3.3&   2.7\\
  13069&  212.6&   49.0&  2.92E-17&  2.91E+42&  2.92E+28& -19.6&   2.6&   4.1\\
  15427&  989.4&  228.2&  2.77E-17&  2.76E+42&  5.96E+27& -17.8&   2.5&   0.8\\
  15823&  333.6&   76.9&  2.73E-17&  2.72E+42&  1.74E+28& -19.0&   2.5&   2.4\\
  17533&  330.0&   76.1&  3.28E-17&  3.27E+42&  2.12E+28& -19.2&   3.0&   3.0\\
  19286&  412.9&   95.2&  3.18E-17&  3.17E+42&  1.64E+28& -18.9&   2.9&   2.3\\
  27691& 1926.8&  444.4&  4.14E-17&  4.13E+42&  4.58E+27& -17.6&   3.8&   0.6\\
  28584&  112.8&   26.0&  2.21E-17&  2.20E+42&  4.17E+28& -20.0&   2.0&   5.8\\
  28996&  254.9&   58.8&  5.38E-17&  5.37E+42&  4.50E+28& -20.0&   4.9&   6.3\\
  29083&  292.7&   67.5&  2.33E-17&  2.32E+42&  1.70E+28& -19.0&   2.1&   2.4\\
  37812&  174.3&   40.2&  2.94E-17&  2.93E+42&  3.60E+28& -19.8&   2.7&   5.0\\
  38785&  182.3&   42.0&  2.75E-17&  2.74E+42&  3.21E+28& -19.7&   2.5&   4.5\\
  41668&  336.0&   77.5&  2.52E-17&  2.52E+42&  1.60E+28& -18.9&   2.3&   2.2\\
  44636&  159.5&   36.8&  2.64E-17&  2.64E+42&  3.53E+28& -19.8&   2.4&   4.9\\
  47969&  520.4&  120.0&  1.27E-16&  1.27E+43&  5.21E+28& -20.2&  11.6&   7.3\\
  49443&  432.4&   99.7&  2.36E-17&  2.35E+42&  1.16E+28& -18.6&   2.1&   1.6\\
  51844&  536.8&  123.8&  3.78E-17&  3.77E+42&  1.50E+28& -18.9&   3.4&   2.1\\
  54027&  181.6&   41.9&  3.42E-17&  3.41E+42&  4.02E+28& -19.9&   3.1&   5.6\\
  55020&  558.4&  128.8&  2.53E-17&  2.52E+42&  9.64E+27& -18.4&   2.3&   1.4\\
  55100&  316.0&   72.9&  2.53E-17&  2.52E+42&  1.71E+28& -19.0&   2.3&   2.4\\
  55848&  984.6&  227.1&  1.04E-16&  1.03E+43&  2.24E+28& -19.3&   9.4&   3.1\\
  57223&  632.3&  145.8&  2.78E-17&  2.77E+42&  9.36E+27& -18.3&   2.5&   1.3\\
  57844&  175.1&   40.4&  2.75E-17&  2.74E+42&  3.34E+28& -19.7&   2.5&   4.7\\
  59074&  689.7&  159.1&  2.36E-17&  2.36E+42&  7.30E+27& -18.1&   2.1&   1.0\\
  59405&  204.7&   47.2&  5.80E-17&  5.78E+42&  6.04E+28& -20.4&   5.3&   8.5\\
  60224&  319.9&   73.8&  3.52E-17&  3.51E+42&  2.34E+28& -19.3&   3.2&   3.3\\
  60455&  154.8&   35.7&  3.17E-17&  3.16E+42&  4.36E+28& -20.0&   2.9&   6.1\\
  60664&  259.9&   59.9&  2.20E-17&  2.19E+42&  1.80E+28& -19.1&   2.0&   2.5\\
  63293&  228.1&   52.6&  2.27E-17&  2.27E+42&  2.12E+28& -19.2&   2.1&   3.0\\
  69085&  478.3&  110.3&  2.12E-17&  2.11E+42&  9.44E+27& -18.3&   1.9&   1.3\\
  70551&  236.5&   54.5&  3.78E-17&  3.77E+42&  3.41E+28& -19.7&   3.4&   4.8\\
  71679&  297.5&   68.6&  5.03E-17&  5.02E+42&  3.61E+28& -19.8&   4.6&   5.0\\
  72094&  107.3&   24.8&  2.09E-17&  2.08E+42&  4.15E+28& -19.9&   1.9&   5.8\\
  72647&  610.5&  140.8&  4.07E-17&  4.05E+42&  1.42E+28& -18.8&   3.7&   2.0\\
  74055&  687.6&  158.6&  2.97E-17&  2.96E+42&  9.21E+27& -18.3&   2.7&   1.3\\
  74299&  695.1&  160.3&  2.50E-17&  2.50E+42&  7.67E+27& -18.1&   2.3&   1.1\\
  75353&  499.8&  115.3&  3.30E-17&  3.29E+42&  1.41E+28& -18.8&   3.0&   2.0\\
  77460&  188.5&   43.5&  2.78E-17&  2.78E+42&  3.15E+28& -19.6&   2.5&   4.4\\
  78162&  152.7&   35.2&  7.24E-17&  7.23E+42&  1.01E+29& -20.9&   6.6&  14.2\\
  80388&  700.2&  161.5&  2.92E-17&  2.91E+42&  8.89E+27& -18.3&   2.6&   1.2\\
  81380&  124.5&   28.7&  2.35E-17&  2.34E+42&  4.02E+28& -19.9&   2.1&   5.6\\
  82145&  290.3&   67.0&  2.58E-17&  2.58E+42&  1.90E+28& -19.1&   2.3&   2.7\\
  82756&  692.1&  159.6&  4.31E-17&  4.30E+42&  1.33E+28& -18.7&   3.9&   1.9\\
  83157&  391.4&   90.3&  9.34E-17&  9.32E+42&  5.09E+28& -20.2&   8.5&   7.1\\
  86220&  240.8&   55.5&  2.34E-17&  2.33E+42&  2.07E+28& -19.2&   2.1&   2.9\\
  87141&  319.6&   73.7&  5.82E-17&  5.80E+42&  3.88E+28& -19.9&   5.3&   5.4\\
  87553&  489.7&  112.9&  5.07E-17&  5.06E+42&  2.21E+28& -19.3&   4.6&   3.1\\
  88785&  170.3&   39.3&  2.41E-17&  2.40E+42&  3.01E+28& -19.6&   2.2&   4.2\\
  89485&  196.1&   45.2&  2.59E-17&  2.58E+42&  2.82E+28& -19.5&   2.4&   3.9\\
  89896& 1722.7&  397.3&  2.95E-17&  2.95E+42&  3.65E+27& -17.3&   2.7&   0.5\\
  90874&  362.4&   83.6&  3.97E-17&  3.96E+42&  2.33E+28& -19.3&   3.6&   3.3\\
  94639&  850.1&  196.1&  3.57E-17&  3.56E+42&  8.95E+27& -18.3&   3.2&   1.3\\
  94941& 1062.3&  245.0&  2.69E-17&  2.68E+42&  5.39E+27& -17.7&   2.4&   0.8\\
  95272&  665.4&  153.4&  2.44E-17&  2.43E+42&  7.80E+27& -18.1&   2.2&   1.1\\
  96625&  227.3&   52.4&  2.90E-17&  2.89E+42&  2.72E+28& -19.5&   2.6&   3.8\\
  99496&  667.9&  154.0&  3.53E-17&  3.52E+42&  1.12E+28& -18.5&   3.2&   1.6\\
 100145&  426.9&   98.4&  3.34E-17&  3.33E+42&  1.67E+28& -19.0&   3.0&   2.3\\
 101147&  379.6&   87.5&  2.28E-17&  2.27E+42&  1.28E+28& -18.7&   2.1&   1.8\\
 102766&  341.3&   78.7&  2.79E-17&  2.78E+42&  1.74E+28& -19.0&   2.5&   2.4\\
\hline
\multicolumn{9}{c}{IA574}\\
\hline
   6580&  124.7&   26.4&  2.99E-17&  3.88E+42&  7.23E+28& -20.6&   3.5&  10.1\\
  10969&  381.4&   80.7&  8.44E-17&  1.09E+43&  6.68E+28& -20.5&  10.0&   9.4\\
  11379&  420.0&   88.9&  2.86E-17&  3.71E+42&  2.06E+28& -19.2&   3.4&   2.9\\
  13034&  796.8&  168.7&  2.57E-17&  3.33E+42&  9.74E+27& -18.4&   3.0&   1.4\\
  21251&  895.4&  189.6&  2.93E-17&  3.80E+42&  9.88E+27& -18.4&   3.5&   1.4\\
  22223&  352.1&   74.6&  2.81E-17&  3.64E+42&  2.41E+28& -19.4&   3.3&   3.4\\
  25092&  311.1&   65.9&  2.59E-17&  3.36E+42&  2.51E+28& -19.4&   3.1&   3.5\\
  34695&  390.0&   82.6&  6.87E-17&  8.90E+42&  5.31E+28& -20.2&   8.1&   7.4\\
  36987&  805.4&  170.5&  3.08E-17&  4.00E+42&  1.16E+28& -18.6&   3.6&   1.6\\
  45242&  388.8&   82.3&  2.84E-17&  3.69E+42&  2.21E+28& -19.3&   3.4&   3.1\\
  46603& 1243.6&  263.3&  1.24E-16&  1.61E+43&  3.00E+28& -19.6&  14.6&   4.2\\
  51420&  247.0&   52.3&  3.62E-17&  4.70E+42&  4.43E+28& -20.0&   4.3&   6.2\\
  53464&  225.6&   47.8&  3.96E-17&  5.13E+42&  5.29E+28& -20.2&   4.7&   7.4\\
  53930&  265.3&   56.2&  5.29E-17&  6.86E+42&  6.02E+28& -20.4&   6.2&   8.4\\
  54185&  307.9&   65.2&  2.71E-17&  3.51E+42&  2.65E+28& -19.5&   3.2&   3.7\\
  55740&  285.8&   60.5&  2.68E-17&  3.48E+42&  2.83E+28& -19.5&   3.2&   4.0\\
  55786&  286.1&   60.6&  2.78E-17&  3.60E+42&  2.93E+28& -19.6&   3.3&   4.1\\
  57174&  505.4&  107.0&  2.88E-17&  3.73E+42&  1.72E+28& -19.0&   3.4&   2.4\\
  58770&  316.0&   66.9&  3.65E-17&  4.73E+42&  3.48E+28& -19.8&   4.3&   4.9\\
  65463&  184.0&   39.0&  2.95E-17&  3.82E+42&  4.84E+28& -20.1&   3.5&   6.8\\
  65839&  152.8&   32.3&  2.65E-17&  3.44E+42&  5.23E+28& -20.2&   3.1&   7.3\\
  66933&  383.0&   81.1&  2.61E-17&  3.38E+42&  2.05E+28& -19.2&   3.1&   2.9\\
  68446&  258.9&   54.8&  2.85E-17&  3.69E+42&  3.32E+28& -19.7&   3.4&   4.6\\
  69065& 1450.1&  307.0&  2.95E-17&  3.82E+42&  6.14E+27& -17.9&   3.5&   0.9\\
  71041&  219.1&   46.4&  5.18E-17&  6.72E+42&  7.14E+28& -20.5&   6.1&  10.0\\
  72707& 1092.7&  231.4&  2.91E-17&  3.78E+42&  8.05E+27& -18.2&   3.4&   1.1\\
  73613&  343.1&   72.6&  7.85E-17&  1.02E+43&  6.90E+28& -20.5&   9.3&   9.7\\
  74326&  135.7&   28.7&  2.47E-17&  3.20E+42&  5.50E+28& -20.3&   2.9&   7.7\\
  75127&  204.0&   43.2&  4.09E-17&  5.30E+42&  6.05E+28& -20.4&   4.8&   8.5\\
  75930&  893.4&  189.2&  4.23E-17&  5.49E+42&  1.43E+28& -18.8&   5.0&   2.0\\
  76201&  364.5&   77.2&  5.06E-17&  6.56E+42&  4.19E+28& -20.0&   6.0&   5.9\\
  77116&  291.9&   61.8&  3.21E-17&  4.17E+42&  3.32E+28& -19.7&   3.8&   4.7\\
  78959& 1189.9&  251.9&  6.17E-17&  8.00E+42&  1.57E+28& -18.9&   7.3&   2.2\\
  79754& 6840.6& 1448.4&  3.61E-17&  4.69E+42&  1.59E+27& -16.4&   4.3&   0.2\\
  82975& 1087.8&  230.3&  2.81E-17&  3.64E+42&  7.80E+27& -18.1&   3.3&   1.1\\
  85436&  769.0&  162.8&  2.86E-17&  3.71E+42&  1.12E+28& -18.5&   3.4&   1.6\\
  90651& 1303.0&  275.9&  3.41E-17&  4.42E+42&  7.91E+27& -18.1&   4.0&   1.1\\
  91865&  255.4&   54.1&  4.47E-17&  5.80E+42&  5.29E+28& -20.2&   5.3&   7.4\\
  96523&  426.4&   90.3&  3.23E-17&  4.20E+42&  2.29E+28& -19.3&   3.8&   3.2\\
  99338&  637.8&  135.1&  2.92E-17&  3.79E+42&  1.38E+28& -18.8&   3.5&   1.9\\
\hline
\multicolumn{9}{c}{IA598}\\
\hline
  10519&  262.0&   53.1&  3.34E-17&  4.94E+42&  4.58E+28& -20.1&   4.5&   6.4\\
  12390&  110.9&   22.5&  2.61E-17&  3.86E+42&  8.46E+28& -20.7&   3.5&  11.8\\
  13299&  214.0&   43.4&  3.27E-17&  4.83E+42&  5.50E+28& -20.3&   4.4&   7.7\\
  27435&  547.0&  110.9&  3.92E-17&  5.80E+42&  2.58E+28& -19.4&   5.3&   3.6\\
  33977&  265.3&   53.8&  2.49E-17&  3.68E+42&  3.37E+28& -19.7&   3.4&   4.7\\
  37912&  181.9&   36.9&  3.17E-17&  4.69E+42&  6.27E+28& -20.4&   4.3&   8.8\\
  53728&  302.9&   61.4&  2.73E-17&  4.03E+42&  3.24E+28& -19.7&   3.7&   4.5\\
  53761&  169.6&   34.4&  3.11E-17&  4.61E+42&  6.60E+28& -20.4&   4.2&   9.2\\
  55275&  388.4&   78.7&  3.42E-17&  5.05E+42&  3.16E+28& -19.7&   4.6&   4.4\\
  58192&  175.6&   35.6&  2.97E-17&  4.39E+42&  6.07E+28& -20.4&   4.0&   8.5\\
  59826& 1162.9&  235.7&  2.93E-17&  4.33E+42&  9.06E+27& -18.3&   3.9&   1.3\\
  60845&  149.3&   30.3&  2.84E-17&  4.20E+42&  6.84E+28& -20.5&   3.8&   9.6\\
  64675&  299.1&   60.6&  4.34E-17&  6.42E+42&  5.22E+28& -20.2&   5.8&   7.3\\
  65983&  383.9&   77.8&  3.20E-17&  4.73E+42&  3.00E+28& -19.6&   4.3&   4.2\\
  80344&  570.6&  115.7&  3.61E-17&  5.34E+42&  2.28E+28& -19.3&   4.9&   3.2\\
  81987&  261.4&   53.0&  2.70E-17&  3.99E+42&  3.71E+28& -19.8&   3.6&   5.2\\
  85081&  381.6&   77.3&  2.85E-17&  4.21E+42&  2.69E+28& -19.5&   3.8&   3.8\\
  87226&  321.7&   65.2&  3.88E-17&  5.74E+42&  4.34E+28& -20.0&   5.2&   6.1\\
  89279&  224.3&   45.5&  8.82E-17&  1.30E+43&  1.41E+29& -21.3&  11.9&  19.8\\
  92776&  578.8&  117.3&  2.48E-17&  3.66E+42&  1.54E+28& -18.9&   3.3&   2.2\\
\hline
\multicolumn{9}{c}{IA624}\\
\hline
  15802& 1150.5&  224.7&  3.76E-17&  6.21E+42&  1.36E+28& -18.7&   5.7&   1.9\\
  16965&  283.3&   55.3&  3.03E-17&  5.00E+42&  4.45E+28& -20.0&   4.5&   6.2\\
  18537&  196.6&   38.4&  2.54E-17&  4.19E+42&  5.38E+28& -20.2&   3.8&   7.5\\
  18866&  398.9&   77.9&  2.32E-17&  3.82E+42&  2.42E+28& -19.4&   3.5&   3.4\\
  26804&  230.9&   45.1&  2.48E-17&  4.10E+42&  4.48E+28& -20.0&   3.7&   6.3\\
  30834&  666.1&  130.1&  2.72E-17&  4.49E+42&  1.70E+28& -19.0&   4.1&   2.4\\
  31226&  125.6&   24.5&  2.84E-17&  4.69E+42&  9.43E+28& -20.8&   4.3&  13.2\\
  33170& 1607.5&  314.0&  3.84E-17&  6.34E+42&  9.96E+27& -18.4&   5.8&   1.4\\
  38991&  128.9&   25.2&  2.88E-17&  4.75E+42&  9.29E+28& -20.8&   4.3&  13.0\\
  40098&  532.4&  104.0&  4.31E-17&  7.11E+42&  3.37E+28& -19.7&   6.5&   4.7\\
  52800&  489.3&   95.6&  2.30E-17&  3.80E+42&  1.96E+28& -19.1&   3.5&   2.7\\
  53168&  465.8&   91.0&  2.28E-17&  3.76E+42&  2.04E+28& -19.2&   3.4&   2.9\\
  56258&  670.8&  131.0&  3.31E-17&  5.45E+42&  2.05E+28& -19.2&   5.0&   2.9\\
  57053&  355.1&   69.4&  2.55E-17&  4.21E+42&  2.99E+28& -19.6&   3.8&   4.2\\
  59547&  178.6&   34.9&  2.73E-17&  4.50E+42&  6.36E+28& -20.4&   4.1&   8.9\\
  60562&  207.4&   40.5&  2.20E-17&  3.64E+42&  4.43E+28& -20.0&   3.3&   6.2\\
  61041&  228.5&   44.6&  2.81E-17&  4.64E+42&  5.12E+28& -20.2&   4.2&   7.2\\
  61778&  459.3&   89.7&  2.39E-17&  3.95E+42&  2.17E+28& -19.2&   3.6&   3.0\\
  62001&  220.5&   43.1&  3.51E-17&  5.79E+42&  6.63E+28& -20.5&   5.3&   9.3\\
  63836&  492.5&   96.2&  6.35E-17&  1.05E+43&  5.37E+28& -20.2&   9.5&   7.5\\
  68957&  116.2&   22.7&  2.34E-17&  3.86E+42&  8.38E+28& -20.7&   3.5&  11.7\\
  69429&  195.6&   38.2&  2.86E-17&  4.71E+42&  6.08E+28& -20.4&   4.3&   8.5\\
  83307&  922.2&  180.1&  5.19E-17&  8.56E+42&  2.34E+28& -19.3&   7.8&   3.3\\
  93077& 1171.9&  228.9&  3.99E-17&  6.58E+42&  1.42E+28& -18.8&   6.0&   2.0\\
  99353&  553.8&  108.2&  4.03E-17&  6.66E+42&  3.03E+28& -19.6&   6.1&   4.2\\
  99689&  571.6&  111.6&  3.58E-17&  5.91E+42&  2.61E+28& -19.4&   5.4&   3.7\\
\hline
\multicolumn{9}{c}{IA651}\\
\hline
  18239&  656.0&  122.7&  4.16E-17&  7.79E+42&  3.13E+28& -19.6&   7.1&   4.4\\
  18481&  597.6&  111.8&  3.90E-17&  7.30E+42&  3.22E+28& -19.7&   6.6&   4.5\\
  18929&  181.8&   34.0&  2.19E-17&  4.10E+42&  5.94E+28& -20.3&   3.7&   8.3\\
  56248&  241.2&   45.1&  3.30E-17&  6.18E+42&  6.75E+28& -20.5&   5.6&   9.4\\
  58954&  459.4&   85.9&  6.68E-17&  1.25E+43&  7.18E+28& -20.6&  11.4&  10.1\\
  60304&  267.7&   50.1&  2.70E-17&  5.06E+42&  4.98E+28& -20.1&   4.6&   7.0\\
  61944&  657.8&  123.0&  2.48E-17&  4.64E+42&  1.86E+28& -19.1&   4.2&   2.6\\
  62565&  340.9&   63.8&  2.82E-17&  5.29E+42&  4.09E+28& -19.9&   4.8&   5.7\\
  65736&  465.6&   87.1&  2.41E-17&  4.51E+42&  2.55E+28& -19.4&   4.1&   3.6\\
  68104&  410.6&   76.8&  2.65E-17&  4.96E+42&  3.18E+28& -19.7&   4.5&   4.5\\
  70056&  254.0&   47.5&  2.48E-17&  4.64E+42&  4.81E+28& -20.1&   4.2&   6.7\\
  87781&  173.1&   32.4&  2.44E-17&  4.57E+42&  6.96E+28& -20.5&   4.2&   9.7\\
  95072& 1327.7&  248.3&  4.26E-17&  7.98E+42&  1.58E+28& -18.9&   7.3&   2.2\\
  96758&  312.4&   58.4&  2.97E-17&  5.56E+42&  4.69E+28& -20.1&   5.1&   6.6\\
  97673&  369.1&   69.0&  3.06E-17&  5.73E+42&  4.09E+28& -19.9&   5.2&   5.7\\
  98943&  160.0&   29.9&  2.32E-17&  4.35E+42&  7.17E+28& -20.5&   4.0&  10.0\\
\hline
\multicolumn{9}{c}{IA679}\\
\hline
   9065&  476.5&   85.4&  2.29E-17&  4.87E+42&  2.81E+28& -19.5&   4.4&   3.9\\
  11843&  268.6&   48.1&  7.76E-17&  1.65E+43&  1.69E+29& -21.5&  15.0&  23.6\\
  16415&  492.9&   88.3&  3.56E-17&  7.56E+42&  4.22E+28& -20.0&   6.9&   5.9\\
  18878&  615.4&  110.3&  2.55E-17&  5.41E+42&  2.42E+28& -19.4&   4.9&   3.4\\
  19202&  606.7&  108.7&  2.97E-17&  6.30E+42&  2.86E+28& -19.6&   5.7&   4.0\\
  24769&  398.4&   71.4&  5.62E-17&  1.19E+43&  8.24E+28& -20.7&  10.9&  11.5\\
  33768&  198.0&   35.5&  2.12E-17&  4.50E+42&  6.25E+28& -20.4&   4.1&   8.8\\
  46786&  265.8&   47.6&  2.48E-17&  5.27E+42&  5.46E+28& -20.2&   4.8&   7.6\\
  47730&  174.4&   31.2&  2.01E-17&  4.27E+42&  6.74E+28& -20.5&   3.9&   9.4\\
  52102&  458.6&   82.2&  3.19E-17&  6.77E+42&  4.06E+28& -19.9&   6.2&   5.7\\
  57794&  583.6&  104.5&  2.60E-17&  5.51E+42&  2.60E+28& -19.4&   5.0&   3.6\\
  60173&  265.6&   47.6&  3.59E-17&  7.61E+42&  7.88E+28& -20.6&   6.9&  11.0\\
  62116&  327.7&   58.7&  3.45E-17&  7.31E+42&  6.14E+28& -20.4&   6.7&   8.6\\
  64479&  273.6&   49.0&  2.20E-17&  4.67E+42&  4.70E+28& -20.1&   4.3&   6.6\\
  65576&  137.0&   24.6&  2.74E-17&  5.80E+42&  1.17E+29& -21.1&   5.3&  16.3\\
  74655&  624.1&  111.8&  2.25E-17&  4.77E+42&  2.10E+28& -19.2&   4.3&   2.9\\
  89098&  469.6&   84.1&  3.14E-17&  6.66E+42&  3.90E+28& -19.9&   6.1&   5.5\\
\hline
\multicolumn{9}{c}{IA709}\\
\hline
   8190&  154.1&   26.5&  2.02E-17&  4.83E+42&  9.00E+28& -20.8&   4.4&  12.6\\
  13683&  978.3&  168.0&  2.59E-17&  6.20E+42&  1.82E+28& -19.1&   5.6&   2.5\\
  18701&  321.4&   55.2&  2.21E-17&  5.30E+42&  4.73E+28& -20.1&   4.8&   6.6\\
  31658&  267.5&   45.9&  2.40E-17&  5.75E+42&  6.17E+28& -20.4&   5.2&   8.6\\
  33815&  177.6&   30.5&  2.78E-17&  6.66E+42&  1.08E+29& -21.0&   6.1&  15.1\\
  34592&  272.0&   46.7&  2.37E-17&  5.68E+42&  6.00E+28& -20.4&   5.2&   8.4\\
  38261&  309.8&   53.2&  2.18E-17&  5.21E+42&  4.83E+28& -20.1&   4.7&   6.8\\
  51167&  225.2&   38.7&  2.28E-17&  5.45E+42&  6.94E+28& -20.5&   5.0&   9.7\\
  55557&  176.5&   30.3&  2.45E-17&  5.88E+42&  9.56E+28& -20.9&   5.3&  13.4\\
  59617&  460.3&   79.0&  6.84E-17&  1.64E+43&  1.02E+29& -20.9&  14.9&  14.3\\
  66451&  171.7&   29.5&  1.96E-17&  4.68E+42&  7.83E+28& -20.6&   4.3&  11.0\\
  72647&  907.7&  155.9&  2.43E-17&  5.83E+42&  1.84E+28& -19.1&   5.3&   2.6\\
  80850&  286.9&   49.3&  2.02E-17&  4.84E+42&  4.84E+28& -20.1&   4.4&   6.8\\
  83599&  946.8&  162.6&  2.81E-17&  6.73E+42&  2.04E+28& -19.2&   6.1&   2.9\\
  84928&  809.3&  139.0&  4.08E-17&  9.76E+42&  3.46E+28& -19.8&   8.9&   4.8\\
\hline
\end{longtable}


\clearpage

\begin{figure}
  \begin{center}
    \FigureFile(80mm,80mm){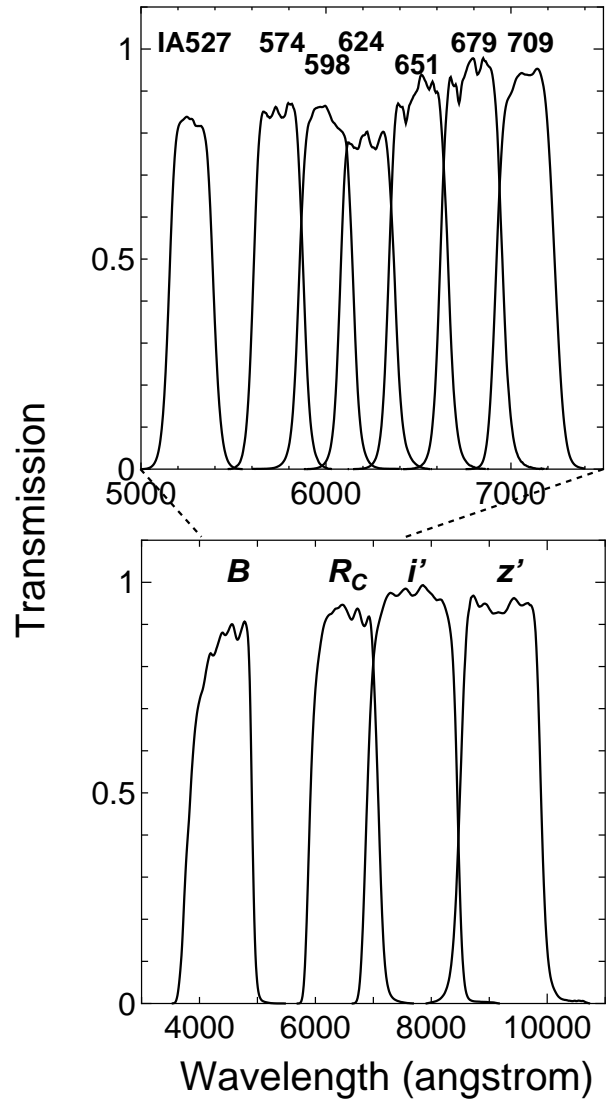}
  \end{center}
  \caption{Transmission curves of the filters used in this study. 
Lower panel shows the four broad band filters and 
upper panel shows a close up for the seven IA filters.
}\label{fig:sample1}
\end{figure}

\begin{figure}
  \begin{center}
    \FigureFile(150mm,150mm){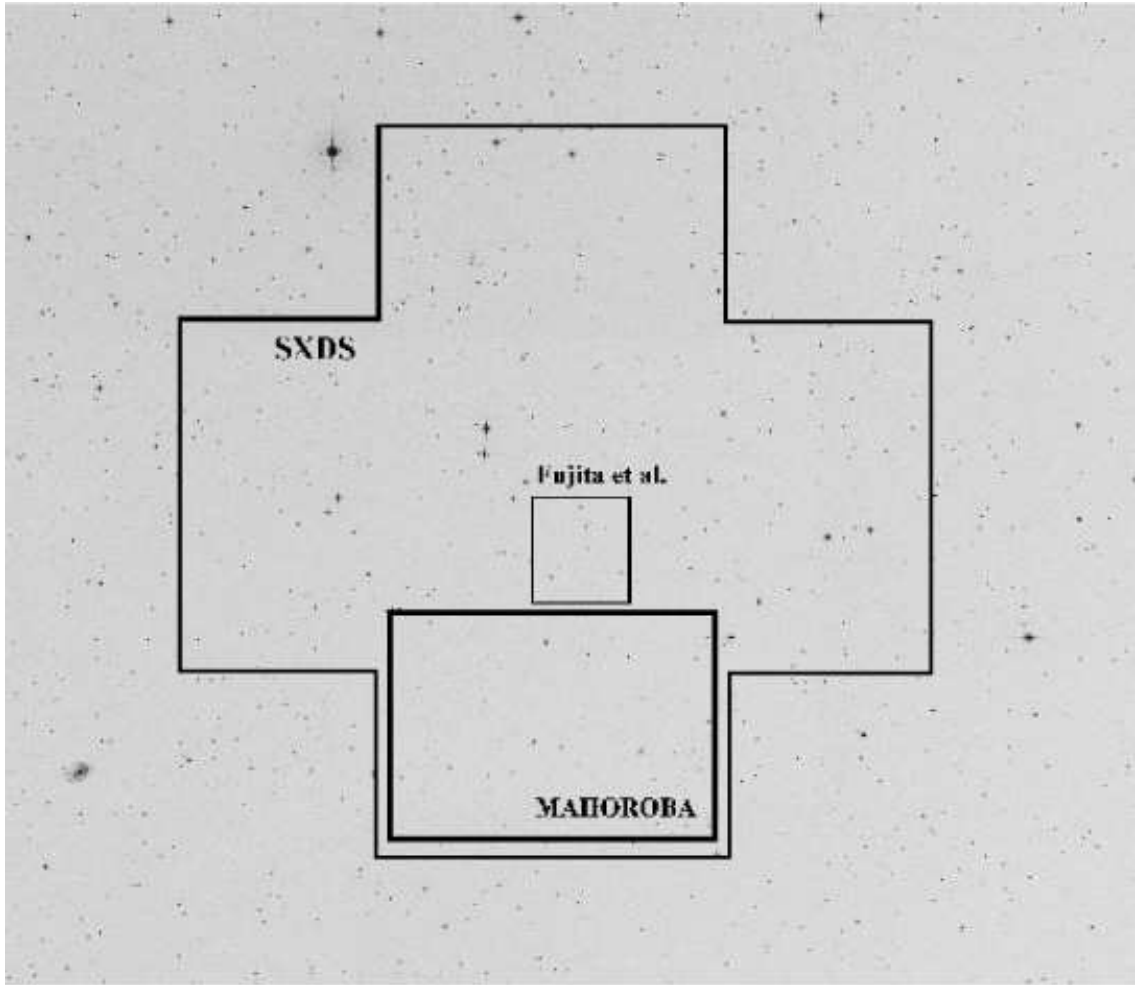}
  \end{center}
  \caption{
The entire filed of view of the SXDS field. Our MAHOROBA field ($34.'71 \times 27.'20$) is
a sky area in the southern part of the SXDS field while
that of Fujita et al. (2003a) is located in the central part($13.'7 \times 13.'7$).
}\label{fig:sample2}
\end{figure}

\begin{figure}
  \begin{center}
    \FigureFile(150mm,150mm){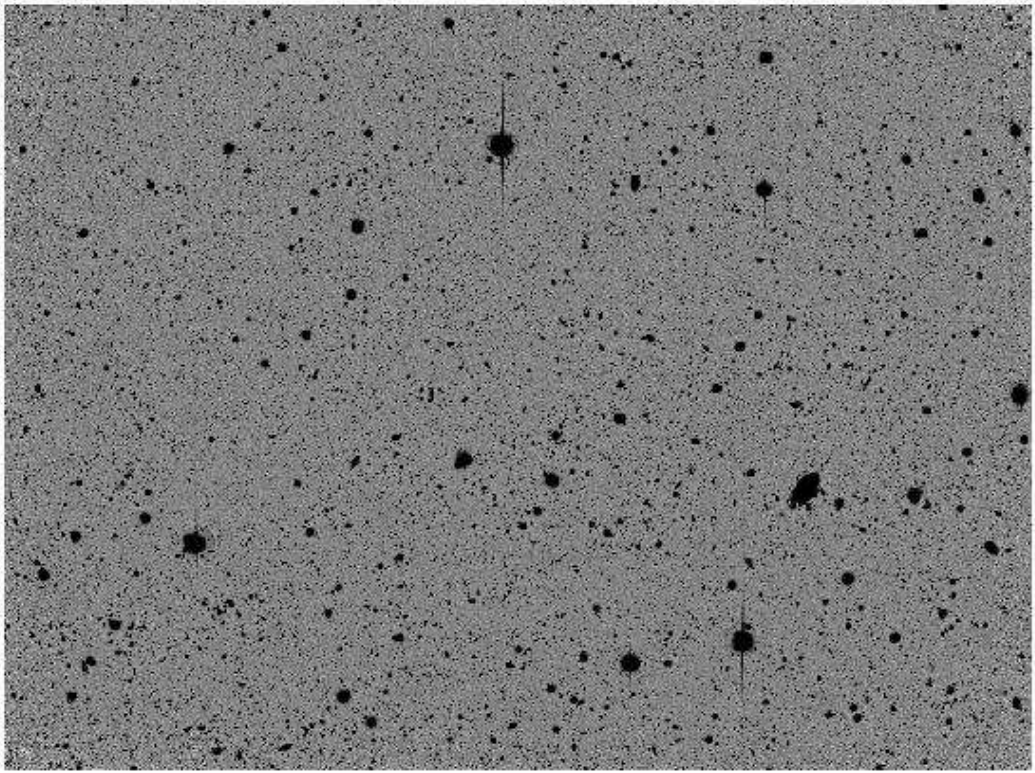}
  \end{center}
  \caption{
The stacked image obtained with the seven IA filters. 
}\label{fig:sample3}
\end{figure}

\begin{figure}
  \begin{center}
    \FigureFile(80mm,80mm){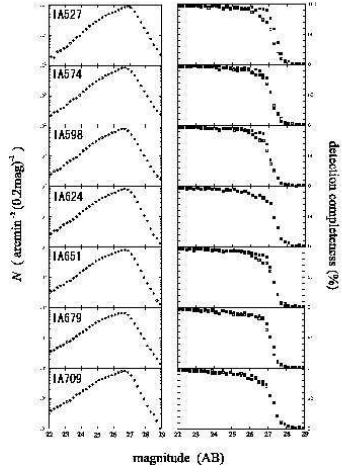}
  \end{center}
  \caption{
Left : Number counts per arcmin$^{\rm 2}$ in 0.2 mag bin. 
Right : The detection completeness of artificial galaxies using the simulation.
Filled squares show the results from the data sets using the exponential law. 
Open squares show that of the de Vaucouleurs law.
}\label{fig:sample4}
\end{figure}

\begin{figure}
    \FigureFile(50mm,50mm){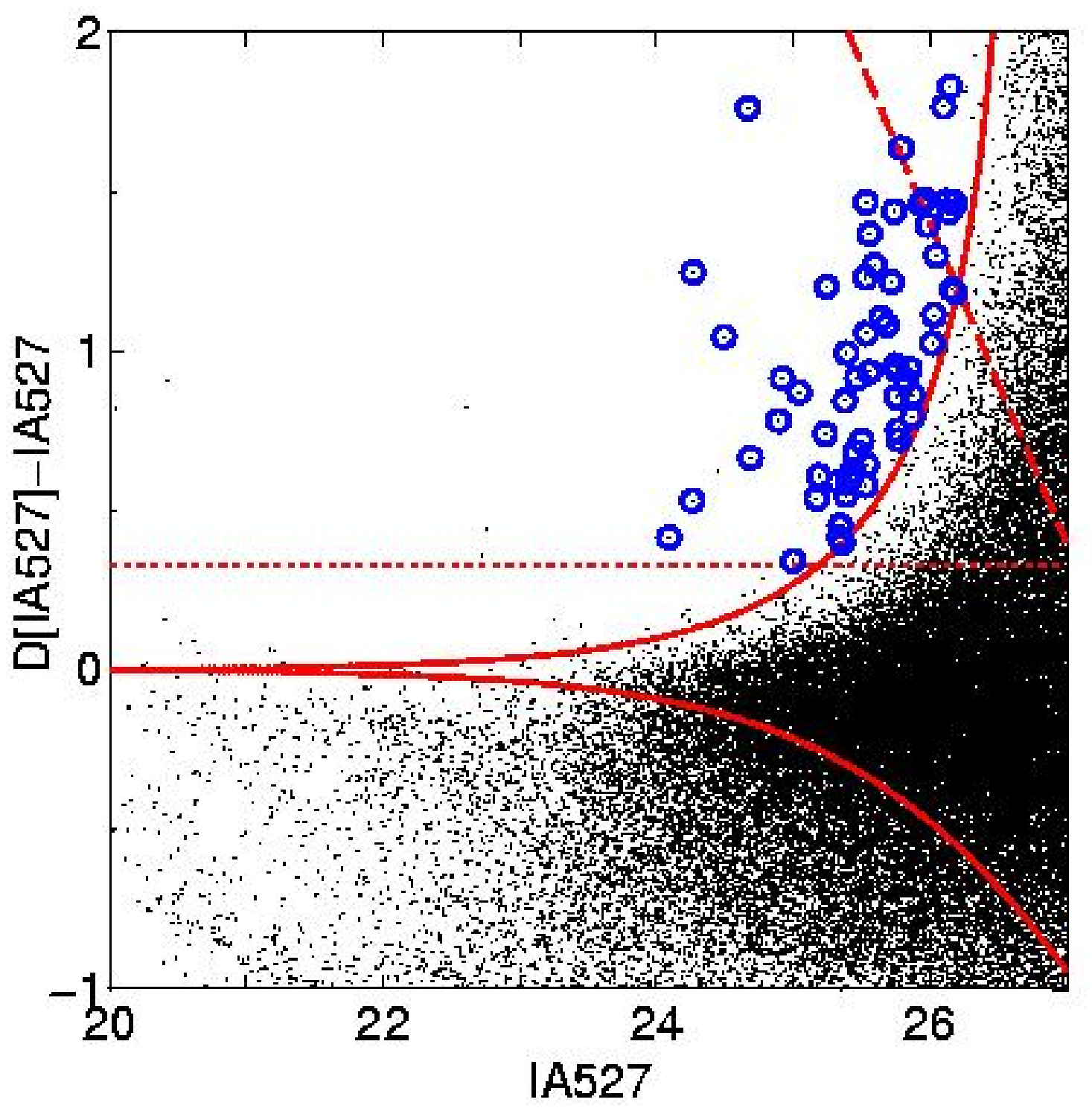}
    \FigureFile(50mm,50mm){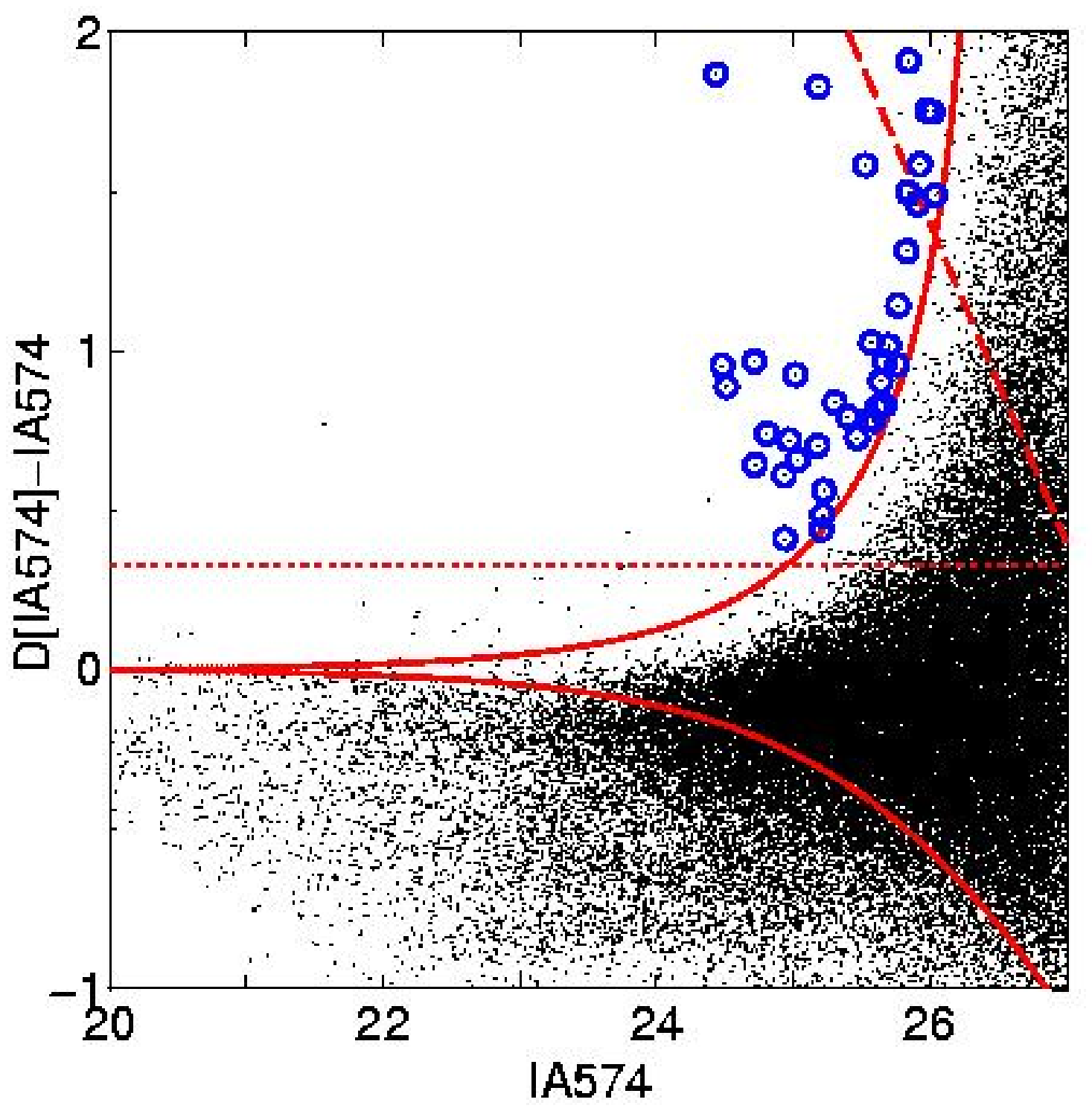}
    \FigureFile(50mm,50mm){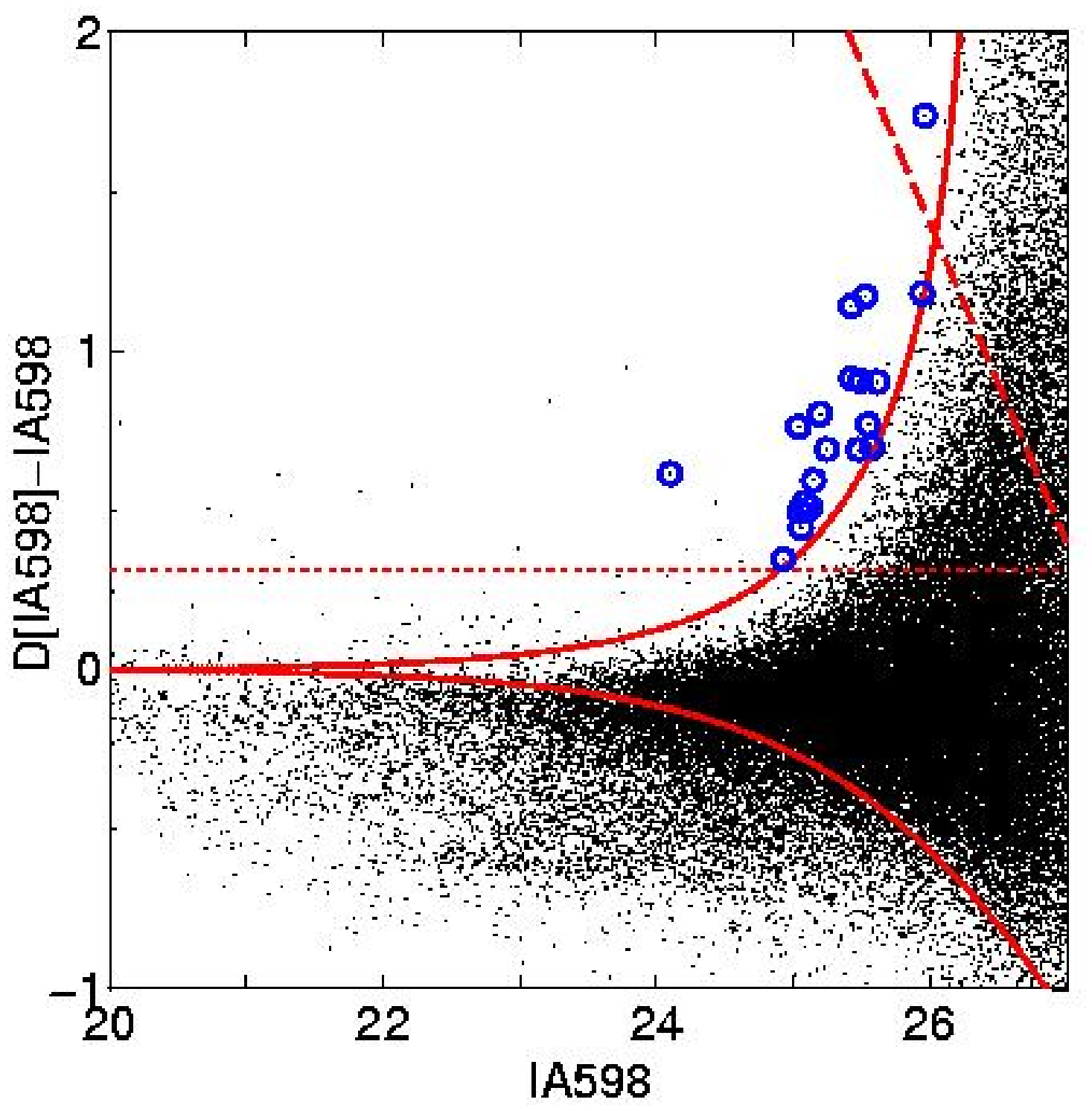}
    \FigureFile(50mm,50mm){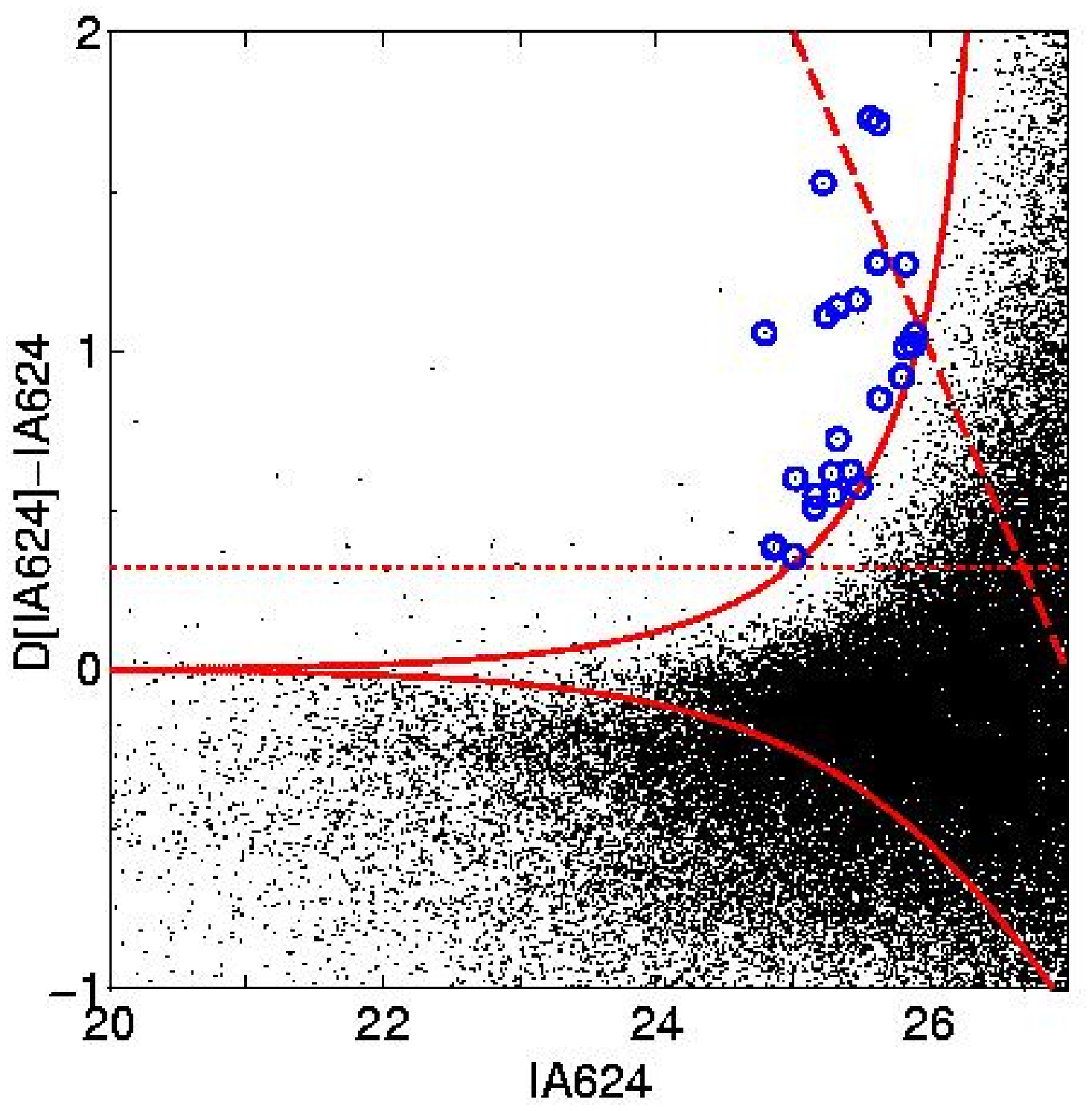}
    \FigureFile(50mm,50mm){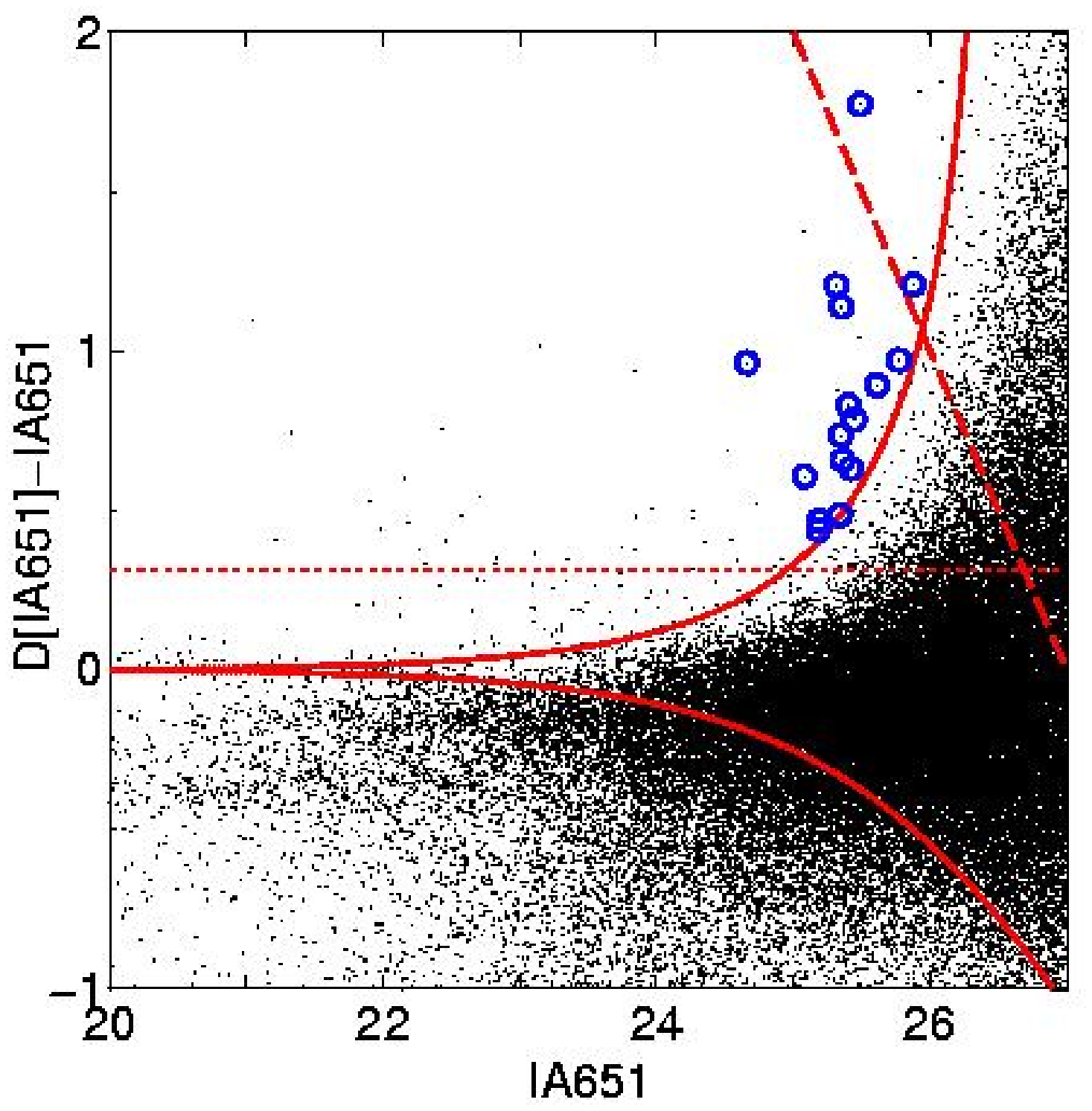}
    \FigureFile(50mm,50mm){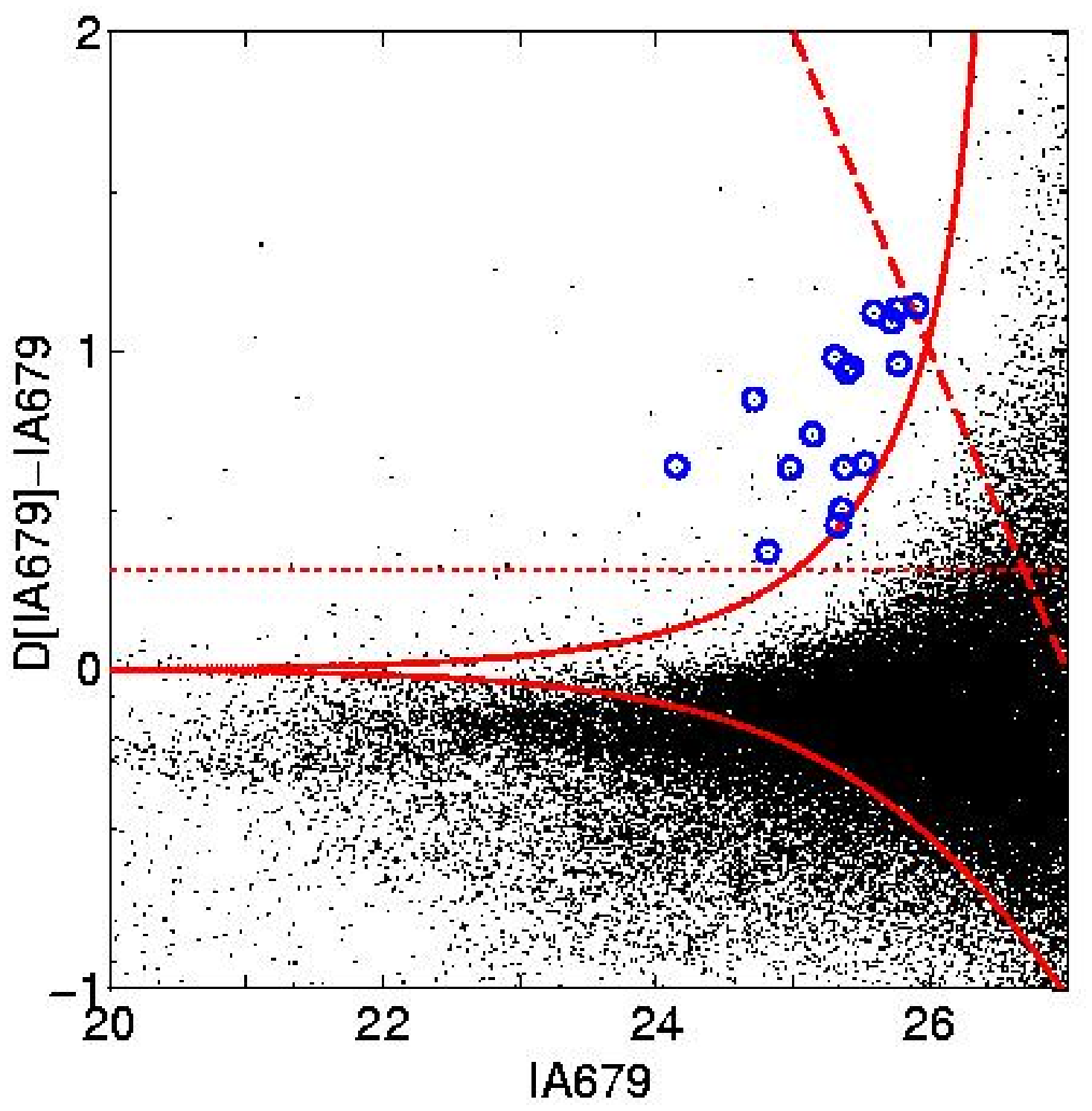}
    \FigureFile(50mm,50mm){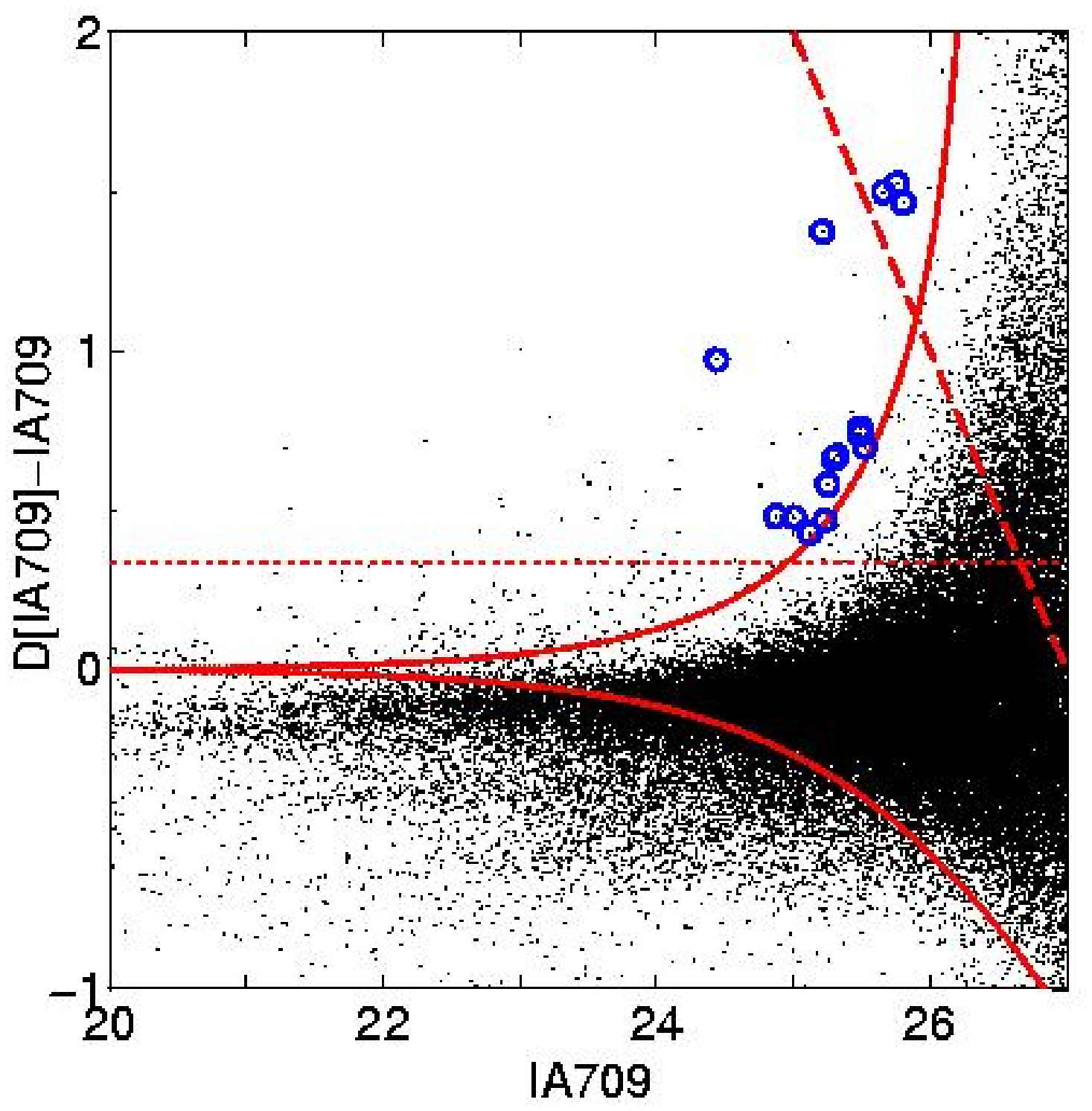}
  \caption{
Color-magnitude diagrams for objects with $IA<$27 at each IA filter. 
Red dotted horizontal line corresponds to the criterion of $EW_{\rm rest}=20$\AA. 
Red dashed line corresponds to 3$\sigma$ of $D[IA]$. 
Red solid curve corresponds to 3$\sigma$ of $D[IA]-IA$. 
Blue open circles show LAE candidates.
}\label{fig:sample5}
\end{figure}

\begin{figure}
    \FigureFile(150mm,150mm){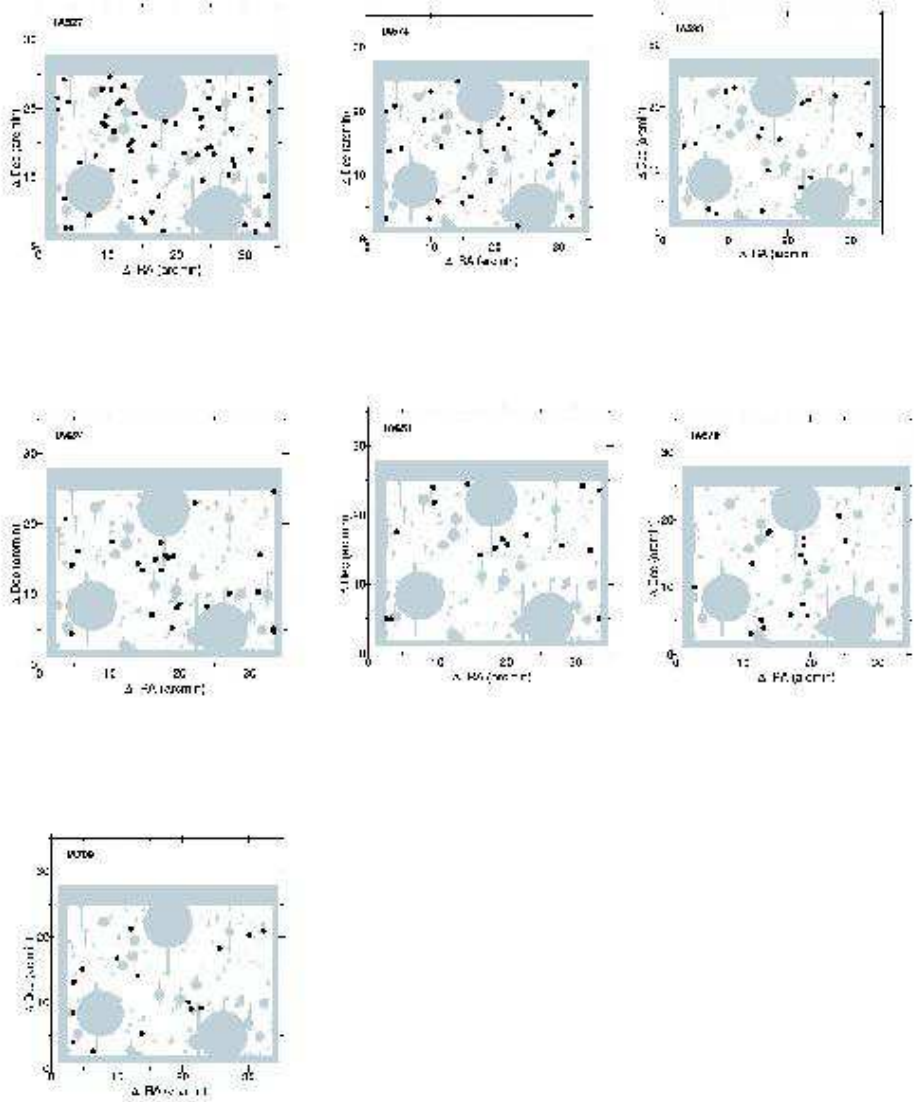}
  \caption{
The spatial distributions of LAE candidates. 
The gray shadow show the masked area. The three large area in these figure 
are masked for the starlight contamination. 
}\label{fig:sample6}
\end{figure}

\begin{figure}
    \FigureFile(150mm,150mm){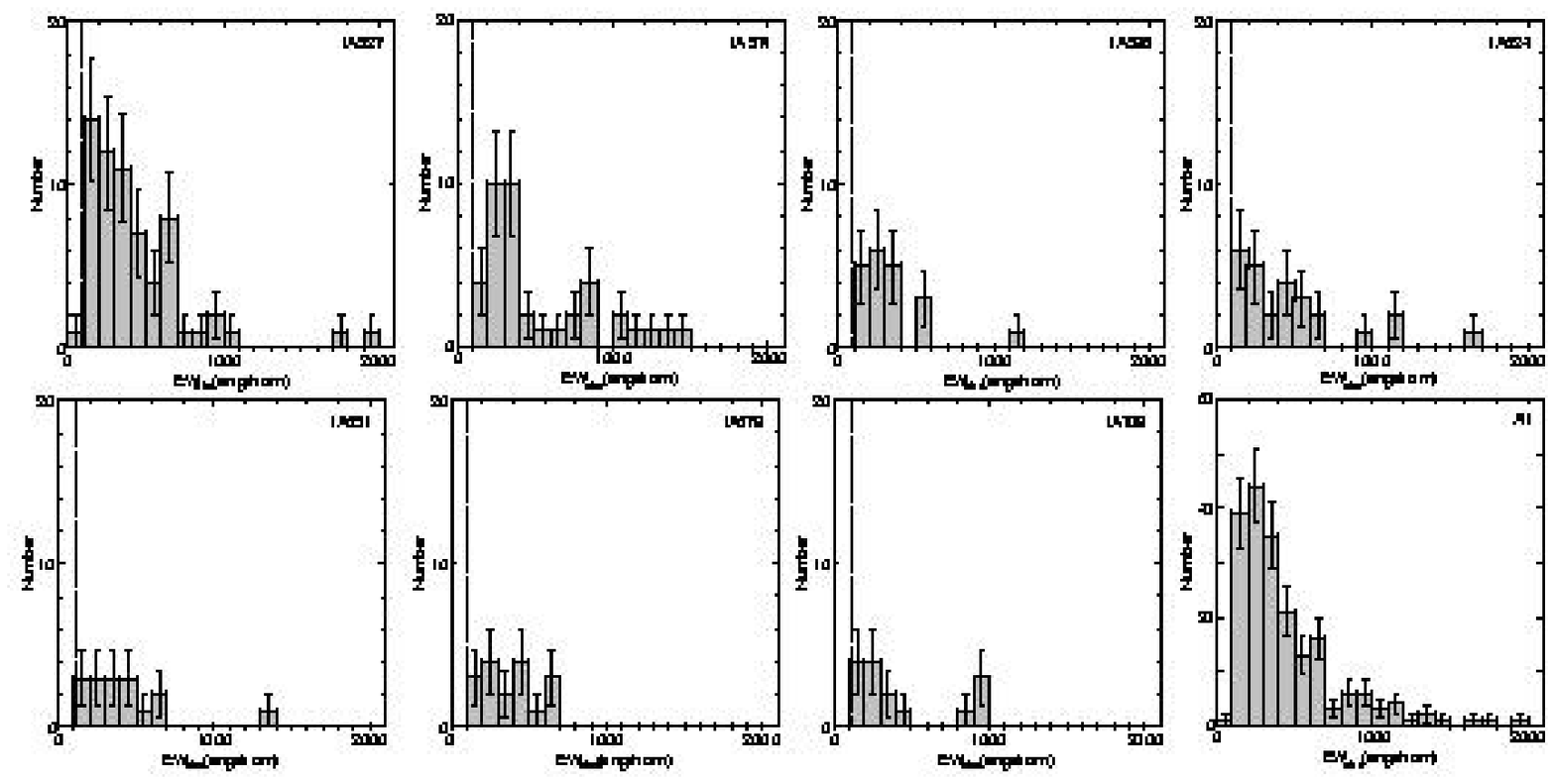}
  \caption{
$EW_{\rm obs}$ distribution of LAE candidates 
found in each IA-excess catalog.
Vertical line in each panel corresponds to $mag(EW_{\rm rest}=20$\AA).
In the panel for $IA574$, one large $EW_{\rm obs}$ 
objects is not shown; its equivalent width is 
$EW_{\rm obs}$ = 6840 \AA. 
}\label{fig:sample7}
\end{figure}

\begin{figure}
    \FigureFile(150mm,150mm){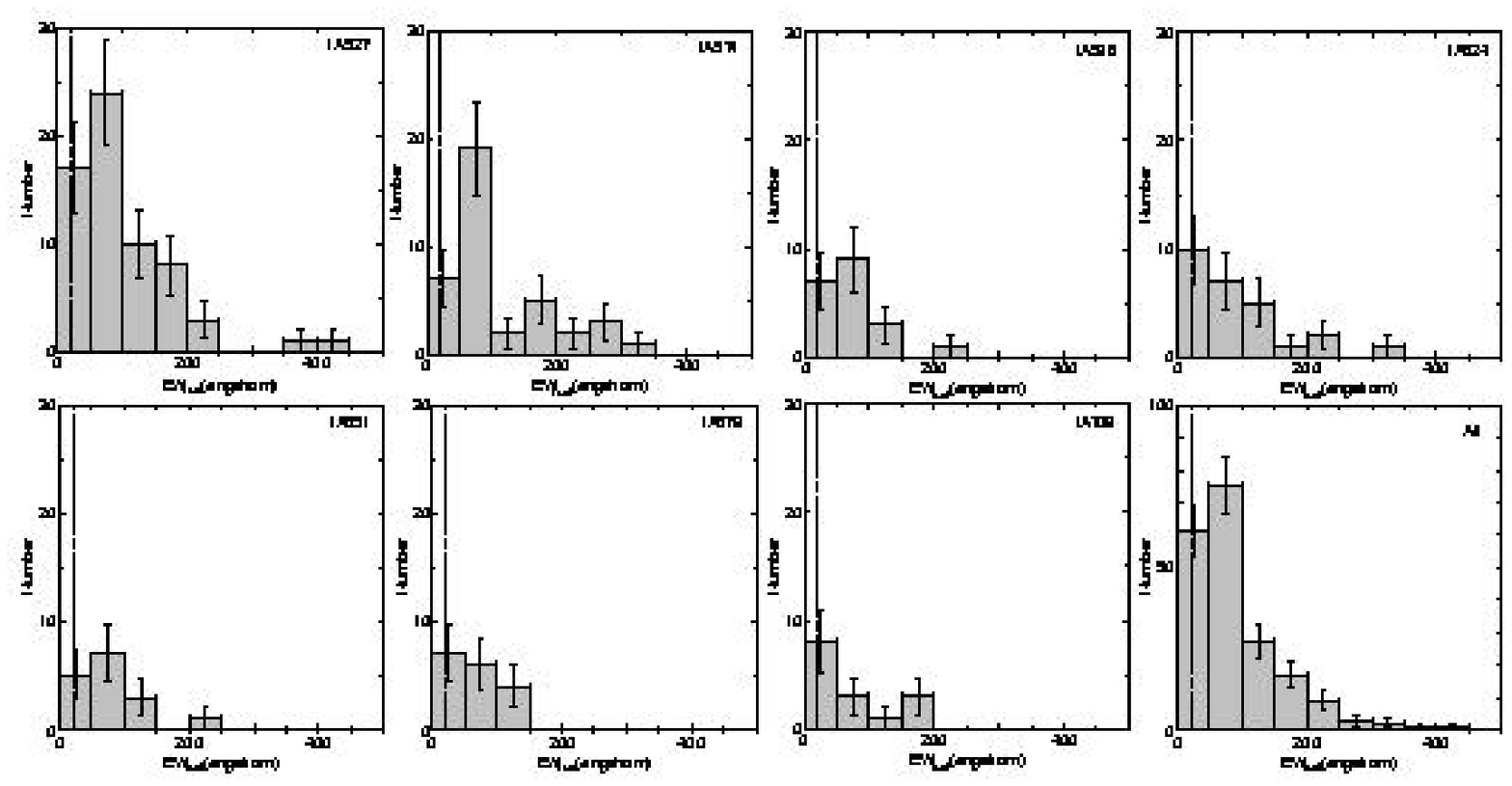}
  \caption{
$EW_{\rm 0}$ distribution of LAE candidates in each IA-excess catalog. 
Vertical line in each panel corresponds to $mag(EW_{\rm rest}=20$\AA).
In the panel for $IA574$, one large $EW_{\rm obs}$ 
objects is not shown; its equivalent width is 
$EW_{\rm obs}$ = 1448 \AA. 
}\label{fig:sample8}
\end{figure}

\begin{figure}
  \begin{center}
    \FigureFile(80mm,80mm){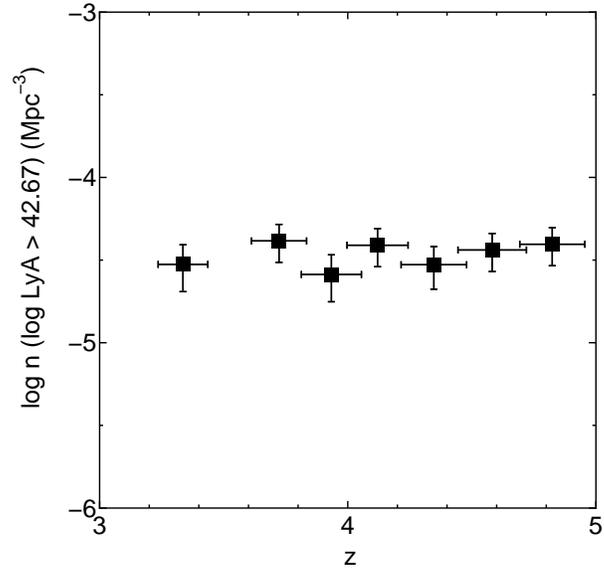}
  \end{center}
  \caption{
Number densities of LAE candidates are shown as a function of
redshift. The horizon error bar simply shows the redshift
coverage of each IA filter. 
}\label{fig:sample9}
\end{figure}

\begin{figure}
    \FigureFile(150mm,150mm){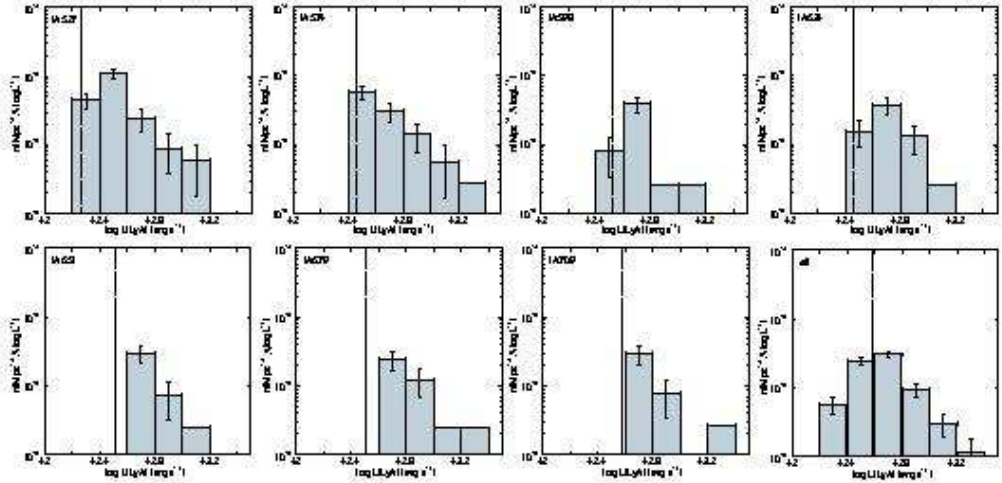}
  \caption{
Luminosity distributions of LAE candidates in each
IA-excess catalog. The lower right panel shows the
results for all LAE candidates. The vertical lines show 
the limit of $L$(Ly$\alpha$) at each filter.
}\label{fig:sample10}
\end{figure}

\begin{figure}
    \FigureFile(150mm,150mm){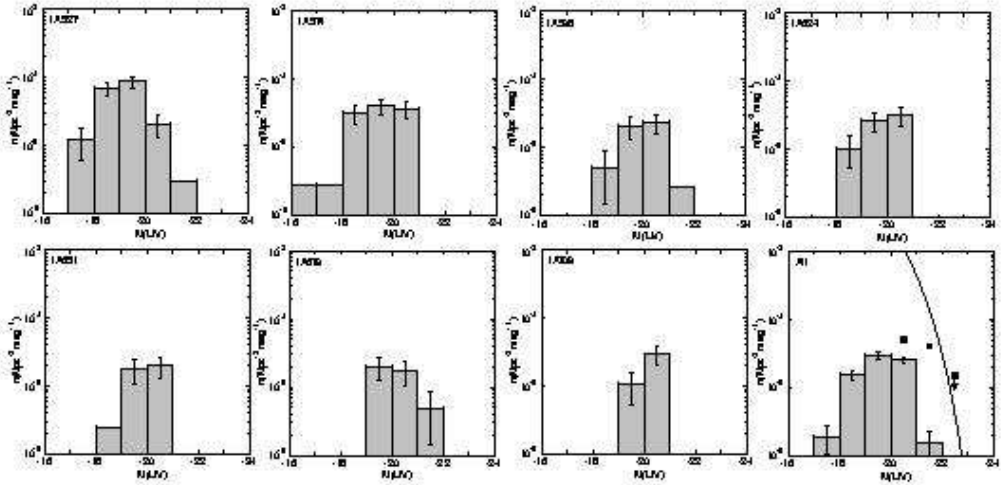}
  \caption{
Luminosity distributions of the absolute $UV$ magnitude for
LAE candidates; 1 $\sigma$ errors are also shown. 
Filled squares in the lower-right panel show the results for $z \sim 5.7$ LAEs 
(Hu et al. 2004). Solid line in the same panel show the UV luminosity function 
for LBGs at $z \sim 4$ (Ouchi et al. 2004). 
}\label{fig:sample11}
\end{figure}

\begin{figure}
    \FigureFile(150mm,150mm){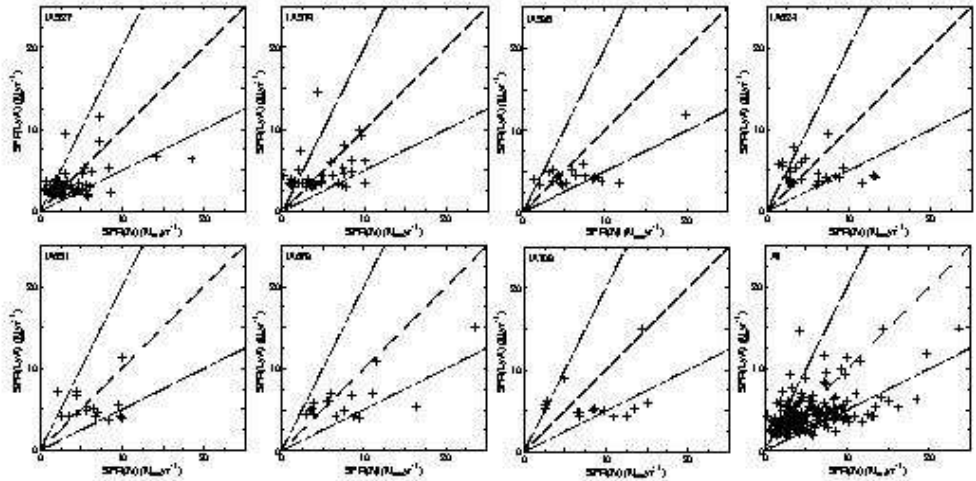}
  \caption{
Comparison between $SFR$(Ly$\alpha$) and $SFR$(UV) for LAEs in each
IA-excess catalog. The following relations are also shown;
1) $SFR$(Ly$\alpha$) = $SFR$(UV) (dashed line),
2) $SFR$(Ly$\alpha$) = $2 \times SFR$(UV) (dotted line),
3) $SFR$(Ly$\alpha$) = $0.5 \times SFR$(UV) (dash-dotted line), and
4) the fitting result (solid line).
The lower-right panel shows the results for all LAEs.
}\label{fig:sample12}
\end{figure}

\begin{figure}
  \begin{center}
    \FigureFile(80mm,80mm){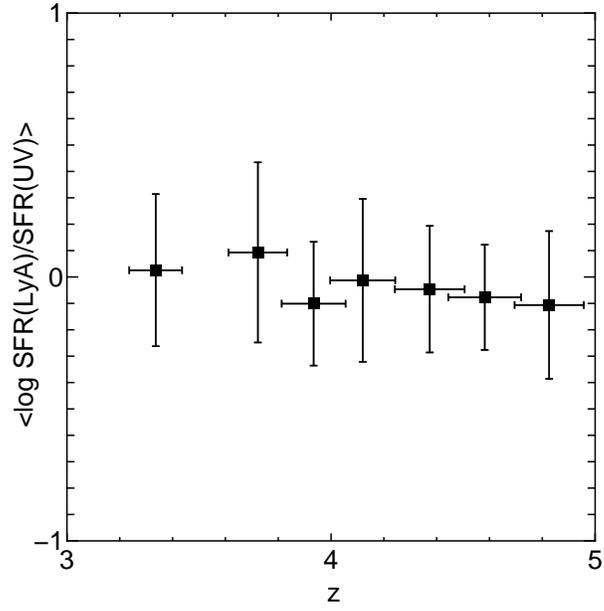}
  \end{center}
  \caption{
The average ratio of $SFR$(Ly$\alpha$) to $SFR$(UV) is shown
As a function of redshift. Our results are shown by filled squares.
}\label{fig:sample13}
\end{figure}

\begin{figure}
  \begin{center}
    \FigureFile(80mm,80mm){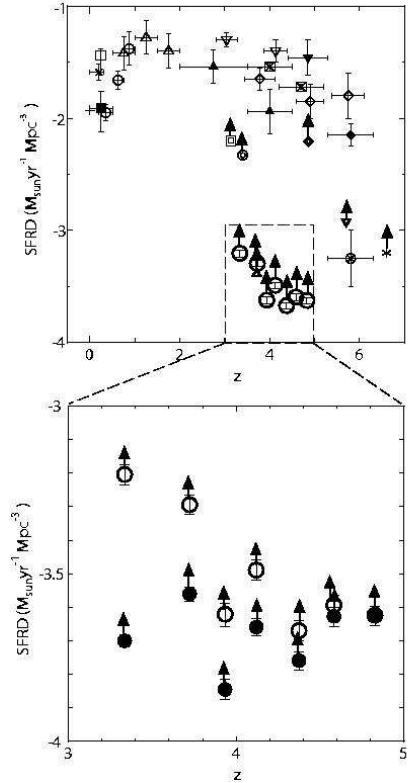}
  \end{center}
  \caption{
Upper panel shows the cosmic evolution of the star formation rate density 
as a function of redshift.
Our results are shown by large open circles (for all LAEs).
Results of previous LAE surveys are also shown;
The data sources are Taniguchi et al. (2005, asterisk), 
Ouchi et al. (2004, double diamond), 
Fujita et al. (2003a, double triangle), 
Cowie \& Hu (1998, double circle), 
Kudritzski et al. (2000, double square). 
All the data above show the lower limit of SFRD. 
We therefore show these data with the up-arrow. 
The data using Lyman break method are shown by the points with error of $SFRD$. 
Those based on LBGs are also shown;
Steidel et al. (2000, open inverse triangles), 
Madau et al. (1998, filled triangles), 
Connolly et al. (1997, open triangles), 
Lilly et al. (1996, open circles),
Iwata et al. (2003, filled inverse triangle),
Giavalisco et al. (2003, open diamonds). 
The other data sources of SFRD are Gallego et al. (1996, plus), 
Tresse \& Maddox (1998, cross), 
Fujita et al. (2003b, open square), 
Treyer et al. (1998, filled square). 

Lower panel shows the cosmic evolution of the star formation rate density 
as a function of redshift. 
We compare the lower limit of SFRD derived from our data. 
The large open circles show the sum of SFRD for all LAEs and 
the large filled circles show the sum of SFRD for LAEs with 
$log ({\rm Ly \alpha}) > 42.67$.
}\label{fig:sample14}
\end{figure}

\begin{figure}
  \begin{center}
    \FigureFile(80mm,80mm){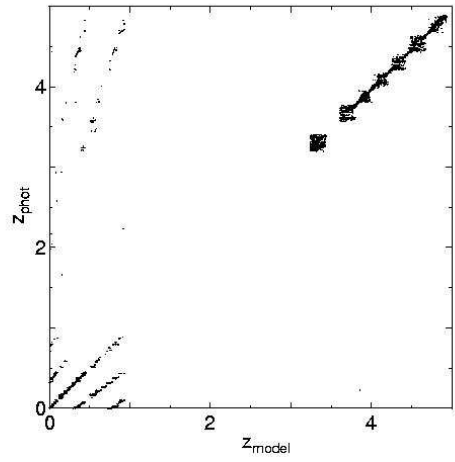}
  \end{center}
  \caption{
Comparison between $z_{\rm model}$ and $z_{\rm phot}$ for simulated catalogs 
selected as emission-line galaxies. 
}\label{fig:sample15}
\end{figure}

\end{document}